\documentclass{vldb}
\usepackage{amsmath}
\usepackage{amsfonts}
\usepackage{amssymb}
\usepackage{mathtools}
\usepackage{graphicx}
\usepackage{balance}
\usepackage{color}
\usepackage{url}
\usepackage{algorithmic}
\usepackage{subfigure}
\usepackage[boxed]{algorithm2e}
\usepackage{multirow}
\usepackage{float}

\newcommand{\QED}{\mbox{\rule[0pt]{1.0ex}{1.0ex}}}
\def\boxend{\hspace*{\fill} $\QED$}

\newtheorem{fact}{Fact}
\newtheorem{lemma}{Lemma}
\newtheorem{theorem}{Theorem}
\newtheorem{corollary}{Corollary}[theorem]

\newcommand{\nop}[1]{}

\DeclareMathOperator*{\argmax}{arg\,max}

\begin{document}
\begin{sloppy}

\title{Activity Maximization by Effective Information Diffusion in Social Networks}

\numberofauthors{1}
\author{
\alignauthor
Zhefeng Wang$^{\dag}$, Yu Yang$^{\ddag}$, Jian Pei$^{\ddag}$ and Enhong Chen$^{\dag}$\\
       \affaddr{$^\dag$University of Science and Technology of China,Hefei, China}\\
        \affaddr{$^\ddag$Simon Fraser University, Burnaby, Canada}\\
       \email{zhefwang@mail.ustc.edu.cn, yya119@sfu.ca, jpei@cs.sfu.ca, cheneh@ustc.edu.cn}
}

\maketitle
\begin{abstract}

In a social network, even about the same information the excitements between different pairs of users are different. If you want to spread a piece of new information and maximize the expected total amount of excitements, which seed users should you choose?  This problem indeed is substantially different from the renowned influence maximization problem and cannot be tackled using the existing approaches.  In this paper, motivated by the demand in a few interesting applications, we model the novel problem of activity maximization.  We tackle the problem systematically. We first analyze the complexity and the approximability of the problem. We develop an upper bound and a lower bound that are submodular so that the Sandwich framework can be applied.  We then devise a polling-based randomized algorithm that guarantees a data dependent approximation factor. Our experiments on three real data sets clearly verify the effectiveness and scalability of our method, as well as the advantage of our method against the other heuristic methods.

\end{abstract}
\section{Introduction}\label{sec:intro}

Based on messages between users in an instant messaging network, such as Whatsapp and WeChat, you can model topics and strengths/frequencies of interaction activities between users. Now you want to raise the awareness of a controversial social issue.  Within a budget, you want to spread the information in the network so that people in the network discuss the issue as much as possible.  Which users should you choose to start spreading the words?

Isn't this an instance of the well known and well studied influence maximization problem~\cite{kempe2003maximizing}?  The answer is ``no'' indeed. Influence maximization selects a seed set of nodes within a given budget constraint such that the expected number of nodes influenced by information diffusion is maximized.  However, to satisfy the requirement that ``people in the network discuss the (target) issue as much as possible'', we not only want to influence many users, but most importantly also want to maximize the expectation of the sum of strength of the interaction activities about the target issue.  This variation of the information diffusion problem, called \emph{activity maximization}, is very different from influence maximization.

Can we adapt some existing influence maximization methods to solve the activity maximization problem? Unfortunately, the answer is no due to the following major reasons. First, the activity maximization problem focuses on the interaction activities between the active users (i.e., the users influenced) while the influence maximization problem aims to simply maximize the expected number of the active users. Since activity strength is different from edge to edge, more active users do not necessarily lead to more interaction activities. Second, at the technical level, the objective functions in the influence maximization problem and the activity maximization problem proposed here have different properties, as to be shown in Section~\ref{sec:properties}. Many existing methods for the influence maximization problem rely on some special properties, such as submodurarity and supermodurarity, of the objective function in influence maximization, which unfortunately do not hold for activity maximization. Third, interaction activities happen between user pairs. The existing methods for influence maximization, however, do not consider activity strengths on edges in their objective functions.

Motivated by the interesting application scenarios and the technical challenges associated, in this paper, we propose a novel problem, activity maximization, which aims to maximize the expectation of the total activity among all active users. A unique novel feature of our problem is that the optimization objective captures interactions among active users. We make several contributions.

First, we identify a novel research problem with interesting applications.  We propose the novel activity maximization problem that aims to maximize the expectation of the overall activities in a social network.  To the best of our knowledge, we are the first to explore the interactions among active nodes in information propagation.

Second, we assess the challenges of the proposed activity maximization problem.  We show that the activity maximization problem is NP-hard under the two most popularly used information diffusion models, namely the independent cascade (IC) model and the linear threshold (LT) model. We also prove that computing the activities with respect to a given set of nodes is \#P-hard under both the IC model and the LT model. Moreover, we show that the objective function of the problem is neither submodular nor supermodular. The theoretical results clearly show that the proposed activity maximization problem cannot be easily solved using the existing methods for influence maximization.  To understand the feasibility of approximate solutions, We appraise the approximability of the problem by constructing a reduction from the densest $k$-subgraph problem.

Third, to develop practical approximate solutions, we develop a lower bound and an upper bound of activities. We prove that maximizing the lower bound or upper bound is still NP-hard under the IC model and the LT model. Moreover, computing the lower bound or upper bound is still \#P-hard under the IC model and the LT model. However, we show the submodularity of the lower bound and the upper bound, which facilitates approximation.

Fourth, we develop a polling based randomized algorithm. We design a sampling method to obtain an unbiased estimation of activities. We also show how to efficiently implement the greedy strategy on the estimate of activities. We extend the sandwich approximation scheme to prove that the proposed algorithm has a data dependent approximation factor.

Last, we verify our algorithm on three real world data sets. The experimental results confirm the effectiveness and the efficiency of the proposed algorithm.

The rest of the pager is organized as follows. We formulate the activity maximization problem in Section~\ref{sec:formulation}. In Section~\ref{sec:properties}, we observe several interesting and useful properties of the proposed problem. We develop a lower bound and an upper bound in Section~\ref{sec:bound}. In Section~\ref{sec:alg}, we devise the polling based algorithm. We review the related work in Section~\ref{sec:rel}. We report the empirical evaluation results in Section~\ref{sec:exp}, and conclude the paper in Section~\ref{sec:con}.  Table~\ref{tab:notation} summarizes the frequently used symbols and their meanings.

\section{Problem Formulation}\label{sec:formulation}
In this section, we first review two widely used information diffusion models, and then give the formal statement of the activity maximization problem.

\begin{table}[t]
\centering\small
\begin{tabular}{|c|p{55mm}|}
\hline
Notation & Description \\ \hline
$G=(V,E,B)$ & A social network, where each edge $(u,v) \in E$ is associated with a diffusion model-dependent parameter $B_{u,v}$\\ \hline
$G_S=(V_S,E_S)$ & The propagation subgraph induced by seed set $S$, where $V_S$ is the set of all active nodes and $E_S=\{(u,v)\mid u\in V_S \land v\in V_S\}$ \\ \hline
$n=|V|$& The number of nodes in $G$ \\ \hline
$A_{u,v}$ & The interaction strength of edge $(u,v)$  \\ \hline
$\delta_A(S)$ & The activity of a given seed set $S$ \\ \hline
$\delta_L(\cdot)$, $\delta_U(\cdot)$ & The lower bound and the upper bound respectively \\ \hline
$g$ & A ``live-edge'' graph instance of $G$\\ \hline
$g\sim G$& $g$ is sampled from all possible instances of $G$ \\ \hline
$R_g(S)$ & The set of nodes reachable from node set $S$ in $g$\\ \hline
$g^T$ & The transpose graph of $g$: $(u,v)\in g\text{ iff }(v,u)\in g^T$ \\ \hline
$R_{g^T}(v)$ & The reverse reachable (RR) set for node $v$\\ \hline
$\mathcal{H}$ & The hypergraph consist of hyperedges\\ \hline
$m_{\mathcal{H}}$ & The number of the hyperedges in $\mathcal{H}$\\ \hline
$\mathcal{D}(S)$ & The degree of the node set $S$ in $\mathcal{H}$\\ \hline

\end{tabular}
\caption{Frequently used notations.}
\label{tab:notation}
\end{table}

\subsection{Diffusion Models}
The independent cascade (IC) model and the linear threshold (LT) model~\cite{kempe2003maximizing} are the two most widely used information diffusion models. Our discussion in this paper is based on these two models. We briefly review them here.

Consider a social network $G=(V,E,B)$, where $V$ is a set of vertices, $E \subseteq V\times V$ is a set of edges, and $B$ is a diffusion model-dependent parameter. Specifically, in the IC model, $B_{u,v}$ is the propagation probability of edge $(u,v)$, which is the probability that $v$ is activated by $u$ after $u$ is activated. In the LT model, $B_{u,v}$ is the influence weight of edge $(u,v)$, which indicates the importance of $u$ influencing $v$.

Both models assume a seed set $S\subseteq V$. Let $S_t$ be the nodes that are activated in step $t$ $(t=0,1,\ldots)$ and $S_0=S$.

In the IC model, the information diffusion process unfolds as follows. At step $t+1$, each node $v$ in $S_t$ has only one chance to activate each inactive neighbor $u$ with the probability $B_{v,u}$. The process terminates when no more nodes can be activated.

In the LT model, the information diffusion process unfolds as follows. Initially, each node $v$ selects a threshold  $\theta_v$ in range $[0,1]$ uniformly at random. At step $t >0$, an inactive node $v$ is activated if $\sum_{w \in N(v)\cap (\mathop{\cup}\limits_{i<t}S_i)}{B_{w,v}} \geq \theta_v$. The process stops at a step $t$ when $S_t=\emptyset$.

Kempe \textit{et al.}~\cite{kempe2003maximizing} also provided an alternative perspective of the information diffusion based on ``live-edge'' graphs. Given a graph $G$, each edge is marked as ``live'' on certain randomized rules, and the random subgraph obtained from all live edges and all nodes in $V$ is called the ``live-edge'' graph~\cite{chen2013information}. Kempe \textit{et al.}~\cite{kempe2003maximizing} proved that we can construct equivalent ``live-edge'' graph models for both IC model and LT model. For the IC model, a ``live-edge'' graph instance can be obtained by marking each edge $(u,v)$ as ``live'' with probability $B_{u,v}$ independently. For the LT model, the corresponding rule is: each node $v$ marks at most one incoming edge $(u,v)$ as ``live'' with probability $1-\sum_{u\in N(v)}{B_{u,v}}$.

\subsection{Activity Maximization}
The activity maximization problem also considers information diffusion in a social network with an extra parameter $A$. Each edge $(u,v) \in E$ is associated with an activity strength $A_{u,v}$, which captures the interaction strength between $u$ and $v$ when they are both active.

Given a social network $G$, an information diffusion model $\mathcal{M}$, and a seed set $S$, the diffusion process forms a propagation induced subgraph $G_S=(V_S,E_S)$, where $V_S$ is the set of all active nodes and $E_S=\{(u,v)\mid u\in V_S \land v\in V_S\}$ is the set of all edges whose two endpoints are both in $V_S$. Then, we can define the \emph{activity} of a given seed set $S$ as
\begin{equation}
\label{eq:ia_def}
    \delta_A(S)=\mathbb{E}\Bigg[\sum_{(u,v)\in E_S}{A_{u,v}}\Bigg]
\end{equation}
where $\mathbb{E}[\cdot]$ is the expectation operator. Since information diffusion is a stochastic process, we take the expectation with respect to all possible diffusion instances. The activity measures the overall interaction strength among the active nodes and thus can reflect the overall strength of the activity caused by the information propagated in the social network.

Now, we can formally define the activity maximization problem as follows. Given a social network $G$, an information diffusion model $\mathcal{M}$, and a budget $k$, find a seed set $S^\ast$ such that
\begin{equation}
\label{eq:iam_def}
    S^\ast=\arg\max\limits_{\begin{array}{c}
         S \subseteq V \\
         |S|=k
    \end{array}}{\delta_A(S)}
\end{equation}

From the definition, we can see the difference between the proposed activity maximization problem and the traditional influence maximization problem. The traditional influence maximization problem aims to maximize the expected number of the active nodes but does not maximize the interactions among them. In contrast, the proposed activity maximization problem focuses on the activity caused by the information diffused and tries to maximize the interaction strength among the active nodes.
\section{Properties of Activity Maximization}\label{sec:properties}
In this section, we first prove the hardness of the activity maximization problem. Then we discuss the properties of the objective function $\delta_A(\cdot)$. Last, we show the approximability of the problem.

\subsection{Hardness Results}

We first assess the hardness of the activity maximization problem.

\begin{theorem}
\label{th:iam}
Activity maximization is NP-hard under the IC model and the LT model.
\end{theorem}
\begin{proof}
We prove by reducing from the set cover problem~\cite{karp1972reducibility}, which is well known in  NP-complete. Given a ground set $\mathcal{U}=\{u_1,u_2,\ldots,u_n\}$ and a collection of sets $\{S_1,S_2,\ldots,S_m\}$ whose union equals the ground set, the set cover problem is to decide if there exist $k$ sets in $\mathcal{S}$ so that the union equals $\mathcal{U}$.

Given an instance of the set cover problem, we construct a corresponding graph with $2n+m$ nodes as follows. We create a node $x_i$ for each set $S_i$, two nodes $y_j$ and $y'_j$ for each element $u_j$, and two edges $(x_i,y_j)$ and $(x_i,y'_j)$ with propagation probability $1$ for the IC model and with influence weight $1$ for the LT model and activity $0$ if $u_j \in S_i$. We also create an edge between $y_j$ and $y'_j$ with propagation probability $0$ and activity $1$ for each element $u_j$. The information diffusion will be a deterministic process, since all propagation probabilities are either $1$ or $0$. Therefore, the set cover problem is equivalent to deciding if there is a set $S$ of $k$ nodes such that $\delta_A(S)=n$. The theorem follows immediately.
\end{proof}

Activity maximization is NP-hard. Then, what is the hardness of computing the  activity with respect to a given seed set $S$?

\begin{theorem}
\label{th:iac}
Given a seed set $S$, computing $\delta_A(S)$ is \#P-hard under the IC model and the LT model.
\end{theorem}
\begin{proof}
We prove by reducing from the influence spread computation problem, which was proved \#P-hard under the IC model and the LT model~\cite{chen2010scalable,chen2010scalablelt}.

Given an instance of the influence spread computation problem, we keep the same graph $G$ and influence diffusion parameters $B$.  We set $A_{u,v}=1$ for any $u,v \in V$ and compute $x_1=\delta_A(S)$ in the graph $G$. Next, we add a new node $v'$ for each node $v$ in the graph $G$ and an edge between $v$ and $v'$ with propagation probability $1$ for the IC model and with influence weight $1$ for the LT model and activity $1$. Now, we obtain a new graph $G'$ and can compute $x_2=\delta_A(S)$ in the graph $G'$. For any newly added node $v'$, the only way to be activated is through its only neighbor $v$. Moreover, a newly added node $v'$ will be activated if its neighbor $v$ is active, since the propagation probability of the newly added edges is $1$. Thus, $x_2-x_1$ is exactly the influence spread in the graph $G$. The theorem follows immediately.
\end{proof}

In \cite{kempe2003maximizing}, Kempe \textit{et al.} introduced the triggering model that generalizes the IC model and the LT model. In the triggering model, each node $v$ independently chooses a subset of its neighbors as its ``triggering set'' according to some distribution. A node will be activated if at least one node of its triggering set is active. We can see that the reduction we construct in the proof of Theorem~\ref{th:iac} still holds for the triggering model. Thus, we have the following result.
\begin{corollary}
Given a seed set $S$, computing $\delta_A(S)$ is \#P-hard in any triggering model $\mathcal{M}$ if computing influence spread is \#P-hard in $\mathcal{M}$.
\end{corollary}

\subsection{Modularity of Objective Functions}

The objective function of influence maximization is submodular under the IC model and the LT model.  Unfortunately, the objective function in activity maximization is no submodular. Moreover, we can show that $\delta_A(\cdot)$ is not supermodular as well.

\begin{figure}[t]
    \centering
    \subfigure[Counter example 1]{\includegraphics[width=0.2\textwidth]{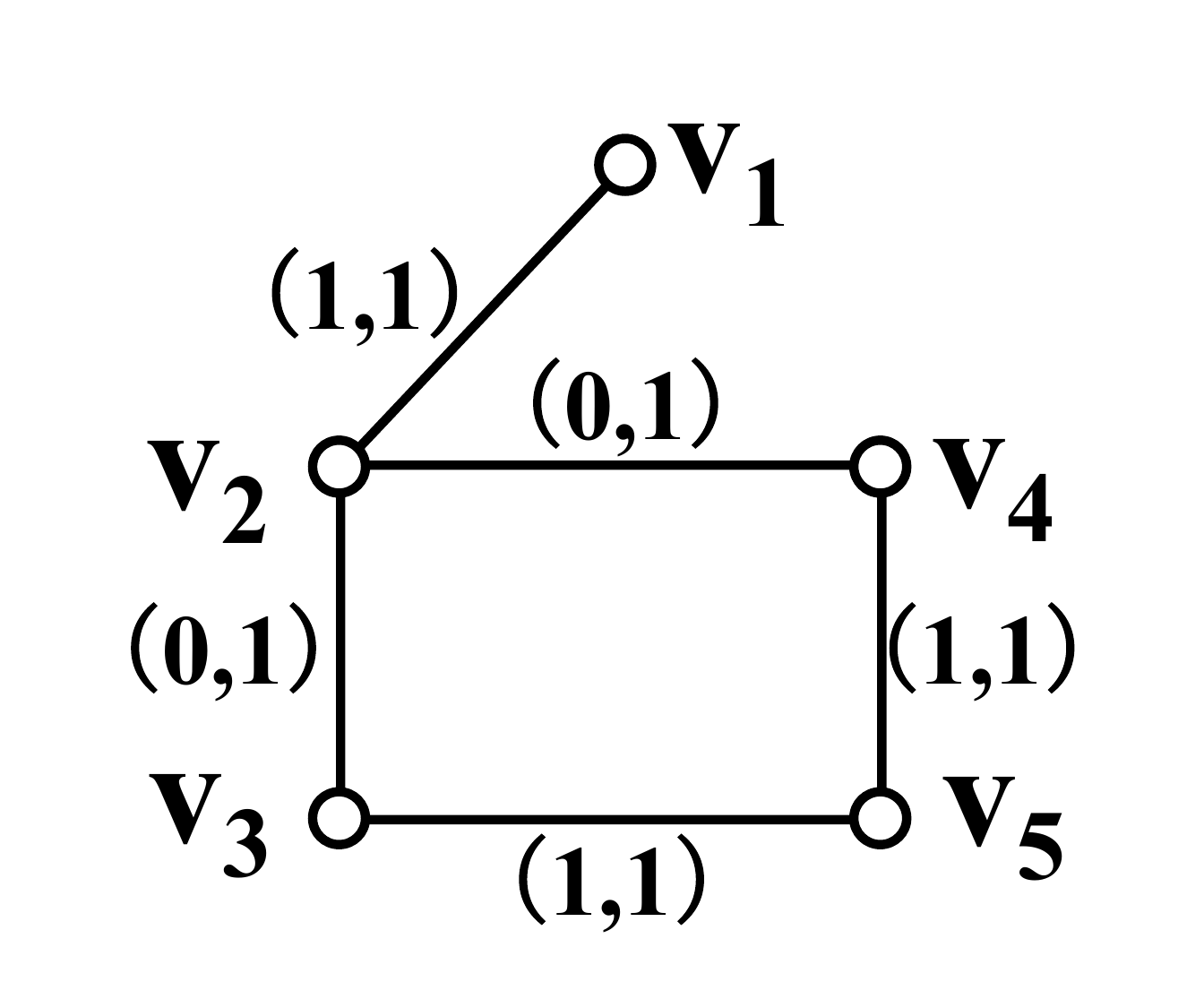}}
    \subfigure[Counter example 2]{\includegraphics[width=0.2\textwidth]{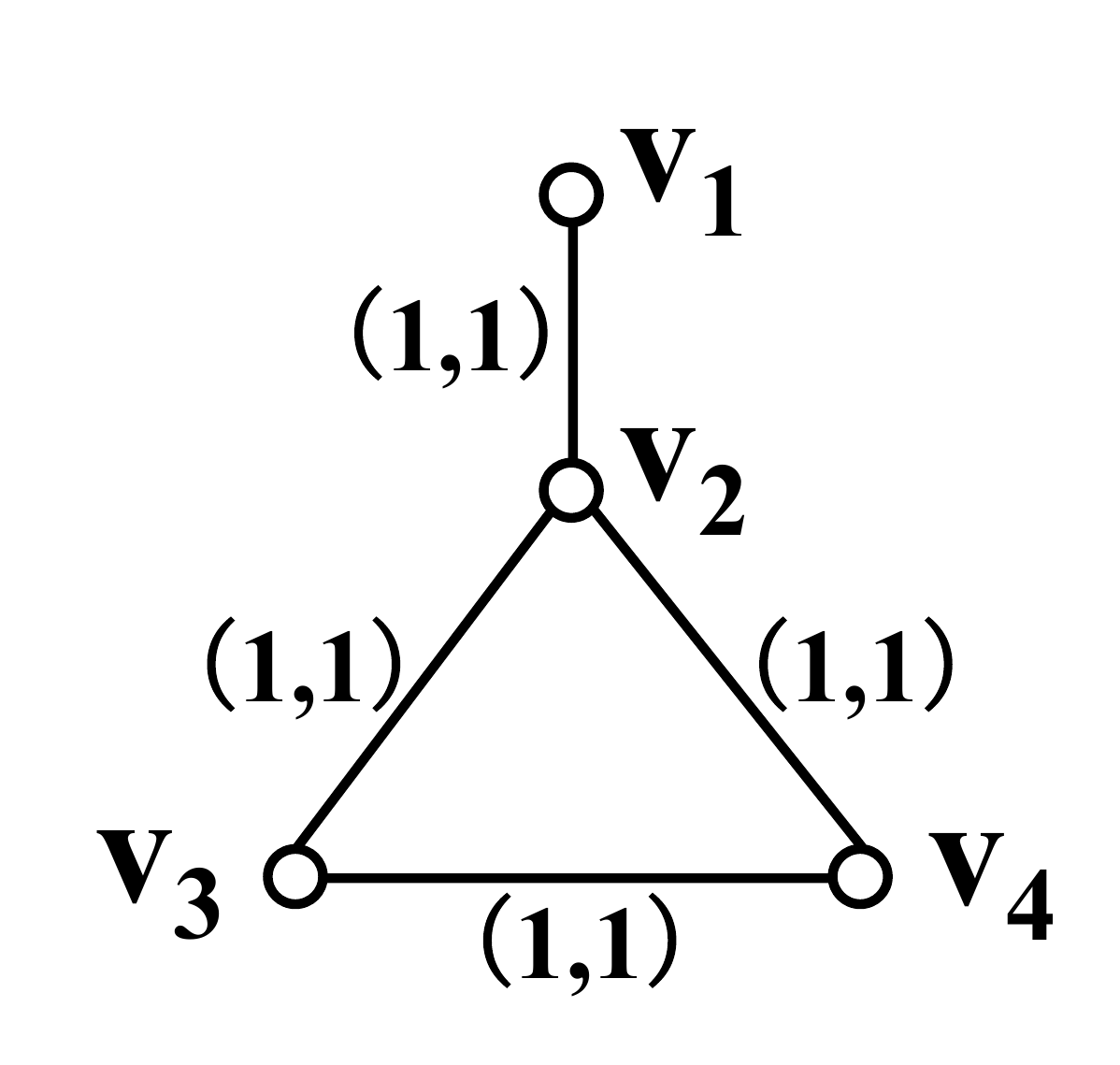}}
    \caption{Counter examples}
    \label{fig:ce}
\end{figure}

\begin{theorem}
$\delta_A(\cdot)$ is not submodular under the IC model and the LT model.
\end{theorem}
\begin{proof}
We prove by a counter example. Consider Fig.~\ref{fig:ce}(a). The first number in the tuple on each edge represents the propagation probability for the IC model and the influence weight for the LT model.  The second number is the activity of the edge. For example, in the counter example 1, $(1,1)$ on edge $(v_1,v_2)$ means $B_{v_1,v_2}=1$ and $A_{v_1,v_2}=1$. In this example, we have $\delta_A(\{v_1\})=1$, $\delta_A(\{v_1,v_5\})=5$ and $\delta_A(\{v_5\})=2$. That is, $$\delta_A(\{v_1\})-\delta_A(\emptyset) < \delta_A(\{v_1,v_5\})-\delta_A(\{v_5\})$$
Therefore, $\delta_A(\cdot)$ is not submodular.
\end{proof}

From counter example 1, we can see that the reason why $\delta_A(\cdot)$ is not submodular is the ``combination effect'' between the newly added node and the existing seed nodes. For example, If we add $v_1$ into $S$ when $S=\emptyset$, then there is only one endpoint is active for edge $(v_2,v_4)$ and $(v_2,v_3)$, that is $v_2$. But if we add $v_1$ to $S$ when $S=\{v_5\}$, then both the two endpoints of edge $(v_2,v_4)$ and $(v_2,v_3)$ are active, since $v_3$ and $v_4$ are activated by $v_5$.

\begin{theorem}
$\delta_A(\cdot)$ is not supermodular under the IC model and the LT model.
\end{theorem}
\begin{proof}
Again, we prove by a counter example. Consider the counter example 2 in Fig~\ref{fig:ce}(b), we have $\delta_A(\{v_2\})=3$, $\delta_A(\{v_1,v_2\})=4$ and $\delta_A(\{v_1\})=4$. Thus, $$\delta_A(\{v_2\})-\delta_A(\emptyset) > \delta_A(\{v_2,v_1\})-\delta_A(\{v_1\})$$  That is, $\delta_A(\cdot)$ is not supermodular.
\end{proof}

\subsection{Approximability}

Since $\delta_A(\cdot)$ is neither submodular nor supermodular, we cannot adopt the standard procedure for optimizing submodular function or supermodular function to get an approximation solution. To explore the approximability of the activity maximization problem, we explore the connection between the activity maximization problem and the densest $k$-subgraph extraction problem.

\begin{theorem}
If there exists a polynomial time algorithm approximating the activity maximization problem within a ratio of $\alpha$, then there exists a polynomial time algorithm that can approximate the densest $k$-subgraph problem within a ratio of $\alpha$.
\end{theorem}
\begin{proof}
We prove by constructing a reduction from the densest $k$-subgraph problem to the activity maximization problem.  Given a graph and an integer $k$, the densest $k$-subgraph problem is to find a subgraph of exactly $k$ vertices that has the maximum density.  For a subgraph $G_S=(V_S,E_S)$, the density is define as $\frac{|E_S|}{|V_S|}$.

Given an instance of the densest $k$-subgraph problem, we construct a corresponding instance of the activity maximization problem. We keep the same graph and set $B_{u,v}=0$ and $A_{u,v}=1$ for $u,v \in V$. Then, the activity maximization problem is to find a set of $k$ vertices and maximize the number of edges whose both endpoints are in this set. It is equivalent to maximizing the density since the number of vertices is constant.
\end{proof}

Khot~\cite{khot2006ruling} showed that the densest $k$-subgraph problem does not admit PTAS\footnote{A PTAS is an algorithm that returns a solution within a factor 1 + $\epsilon$ of being optimal (or 1 - $\epsilon$ for maximization problems) in polynomial time for any $\epsilon>0$.} (Polynomial Time Approximation Scheme~\cite{vazirani2013approximation}) assuming $NP \nsubseteq \mathop{\bigcap}\limits_{\epsilon >0}BPTIME(2^{n^{\epsilon}})$, we immediately have the following result.

\begin{corollary}
There is no PTAS for the activity maximization problem assuming $NP \nsubseteq \mathop{\bigcap}\limits_{\epsilon >0}BPTIME(2^{n^{\epsilon}})$.
\end{corollary}

In fact, finding a good approximation to the densest $k$-subgraph problem is challenging. The current best approximation ratio of $n^{1/4+\epsilon}$ for $\epsilon>0$ was achieved by Bhaskara \textit{et al.}~\cite{bhaskara2010detecting}. It is still unknown if there exists a polynomial algorithm that can approximate the densest $k$-subgraph problem with a constant factor. \nop{Therefore, it is probably hard to approximate the activity maximization problem with a constant factor.}
\section{Lower Bound and Upper Bound}\label{sec:bound}
In this section, we first give a lower bound and an upper bound on activities. Then we discuss the properties of the lower bound and the upper bound.

\subsection{The Bounds}

Since the ``combination effect'' among seed nodes comprises the submodularity of the objective function $\delta_A(\cdot)$, we try to develop a lower bound of $\delta_A(\cdot)$ that is submodular by ignoring the ``combination effect''. The major idea is that we only consider the edges whose two endpoints are activated by the same seed node. Accordingly, the lower bound can be defined as
\begin{equation}
\label{eq:lb_def}
\delta_L(S)=\mathbb{E}[\sum_{(u,v)\in \mathop{\bigcup}\limits_{ x \in S}E_{\{x\}}}{A_{u,v}}]
\end{equation}
where $E_{\{x\}}$ is the set of edges of the propagation subgraph induced by seed set $\{x\}$. Recall that the propagation subgraph induced by a seed set consists of the nodes that can be activated by the seed set. Here, the seed set consists of only one node $x$.
It is easy to see that $\delta_L(S) \leq \delta_A(S)$ for any $S \subseteq V$, since we ignore the edges whose endpoints are activated by different seed nodes.

A straightforward way to get an upper bound is to consider all the edges that have at least one active endpoint. In this way, the upper bound equals to the activity of edges that have one active endpoint plus the activity of edges whose two endpoints are both active. The latter is exactly the activity we want to compute. Here, we present a tighter upper bound from the perspective of active nodes, which can be defined as
\begin{equation}
\label{eq:ub_def}
 \delta_U(S)=\mathbb{E}[\sum_{v\in V_S}{w(v)}]
\end{equation}
where $$w(v)=\frac{1}{2}\sum_{u\in N(v)}{A_{u,v}}.$$ Given a seed set $S$, $\delta_U(S)$ equals to the half of the activity of edges that have one active endpoint plus the activity of edges whose two endpoints are both active. Thus, $\delta_U(S)$ is better than the straightforward one. Also, we can see that the upper bound is essentially a weighted version of the influence spread, where the weight of node $v$ is $\frac{1}{2}\sum_{u\in N(v)}{A_{u,v}}$. For the influence spread, $w(v)=1$ for each node $v$.

\subsection{Properties of the Bounds}

Using the lower bound and the upper bound, we can approximate the information activity problem by maximizing the lower bound and the upper bound~\cite{lu2015competition}. However, maximizing the lower bound and the upper bound is still NP-hard.

\begin{theorem}
Maximizing the lower bound is NP-hard under the IC model and the LT model.
\end{theorem}
\begin{proof}
We prove by reducing from the NP-complete set cover problem~\cite{karp1972reducibility}. We show the reduction constructed in the proof of Theorem~\ref{th:iam} still holds for the lower bound. The lower bound only considers the edges whose two endpoints can be activated by the same seed node. In the previous reduction, for all the edges whose activity is not equal to $0$ (the edges between $y_j$ and $y'_j$), their two endpoints can be activated by the same node. Thus, the set cover problem can be solved by deciding if there is a set $S$ of $k$ nodes such that $\delta_L(S)=n$.
\end{proof}

\begin{theorem}
\label{th:upm}
Maximizing the upper bound is NP-hard under the IC model and the LT model.
\end{theorem}
\begin{proof}
We prove by reducing from the NP-hard influence maximization problem~\cite{kempe2003maximizing}.

Given an instance of the influence maximization problem, let $d_{max}$ be the highest degree of the nodes in the graph $G$. Then, for each node $v$ in $G$, we add $N_d=d_{max}-d_v$ new nodes, $v'_1,v'_2,\ldots,v'_{N_d}$, and  $N_d$ new edges, $(v,v'_1),(v,v'_2),\ldots,(v,v'_{N_d})$. Now we obtain a new graph $G'$. We set the propagation probability  of the newly added edges to $0$ for the IC model, and set the influence weight of the newly added edges to $0$ for the LT model, and set the information activity of all the edges in $G'$ to $\frac{2}{d_{max}}$.

Then, we have $\forall v\in V$, $w(v)=1$,  and $ \forall v'\in V'\setminus V$, $w(v')=\frac{2}{d_{max}}$. Since the propagation probability of all newly added edges is $0$, the newly added nodes can never be activated. Therefore, we have $I^G(S)=\delta_U^{G'}(S),\forall S\subseteq V$, where $I(S)$ is the influence spread of a give seed set $S$ in $G$ and $\delta_U^{G'}(S)$ is the upper bound in $G'$.

Next, we prove that $S_U^\ast=\argmax{\delta_U^{G'}(S)}$ does not contain any newly added nodes. If there is any newly added node in $S_U^\ast$, we can always replace it with a node in $V \setminus S_U^\ast$ and increase the value of the objective function. Thus, if $S_U^\ast$ is the optimal solution of maximizing the upper bound in $G'$, it must be the optimal solution of the influence maximization in $G$. 
\end{proof}

Although maximizing the lower bound and the upper bound is NP-hard, the objective functions of the lower bound and the upper bound are submodular.

\begin{theorem}
\label{th:lbs}
$\delta_L(\cdot)$ is submodular under the IC model and the LT model.
\end{theorem}
\begin{proof}
Given a graph $G$ and an influence diffusion model, either the IC model or the LT model, we can construct ``live-edge'' graphs for $G$ using the methods proposed in~\cite{kempe2003maximizing}. Let $g$ be a ``live-edge'' graph instance. Denote by $Pr(g)$ the probability that $g$ is selected from all possible instances. Let $E_g(S)$ be the set of edges whose two endpoints can be reachable from the same node in the seed set $S$. Then we can rewrite $\delta_L(S)$ to
$$
\delta_L(S)=\sum_{g\sim G}{Pr(g)\sum\limits_{(u,v)\in E_g(S)}{A_{u,v}}}
$$
We only need to prove $Q(S)=\sum\limits_{(u,v)\in E_g(S)}{A_{u,v}}$ is submodular for any ``live-edge'' graph instance $g$, since an non-negative linear combination of submodular functions is also submodular.

To prove, let $M$ and $N$ be two sets such that $M\subseteq N \subseteq V$. For any $v \in V\setminus N$, consider the difference between $Q(M\cup \{v\})$ and $Q(M)$. It must be contributed from the edges whose two endpoints can be reachable from $v$ but cannot be reachable from the nodes in $M$. These edges must be a super set of the edges whose two endpoints can be reachable from $v$ but cannot be reachable from the nodes in $N$, since $M\subseteq N$. It follows that $Q(M\cup \{v\})-Q(M) \geq Q(N\cup \{v\})-Q(N)$. Therefore, $Q(S)$ is submodular and the theorem follows.
\end{proof}

\begin{theorem}\label{th:ubs}
$\delta_U(\cdot)$ is submodular under the IC model and the LT model.
\end{theorem}
\begin{proof}
We can prove the theorem by the same ``live-edge'' technique used in the proof of Theorem~\ref{th:lbs}. Let $R_g(S)$ be the set of nodes reachable from $S$ in $g$. Then, $\delta_U(S)$ can be rewritten to
$$
\delta_U(S)=\sum_{g\sim G}{Pr(g)\sum\limits_{v\in R_g(S)}{w(v)}}
$$

The way to prove that $Q'(S)=\sum\limits_{v\in R_g(S)}{w(v)}$ is submodular is similar to the proof of $Q(S)$ in Theorem~\ref{th:lbs}. The nodes that can be reachable from $v$ but cannot be reachable from the nodes in $M$ must be a super set of the nodes that can be reachable from $v$ but cannot be reachable from the nodes in $N$. It follows that $Q'(M\cup \{v\})-Q'(M) \geq Q'(N\cup \{v\})-Q'(N)$. Therefore, $Q'(S)$ is submodular and the theorem follows.
\end{proof}

Theorems~\ref{th:lbs} and~\ref{th:ubs} are good news. With the submodularity we can adopt the standard procedure for optimizing submodular functions to obtain an approximation solution~\cite{nemhauser1978analysis}. One challenge remains.  Applying the algorithm proposed in~\cite{nemhauser1978analysis} requires evaluating the lower bound and the upper bound. However, computing the lower bound and the upper bound with respect to a given seed set is unfortunately \#P-hard.

\begin{theorem}
Given a seed set $S$, computing $\delta_L(S)$ is \#P-hard under the IC and the LT model.
\end{theorem}
\begin{proof}
We prove by reducing from the influence spread computation problem. We show that the reduction we construct in the proof of Theorem~\ref{th:iac} still holds for the lower bound case. Let $y_1=\delta_L(S)$ in the graph $G$ and $y_2=\delta_L(S)$ in the graph $G'$. Since the propagation probability of the edge $(v,v')$ is $1$ for the IC model and the influence weight of the edge $(v,v')$ is $1$ for the LT model, $v$ and $v'$ can be activated by the same seed node. It follows that $y_2-y_1$ is also the influence spread in the graph $G$.
\end{proof}
\begin{theorem}
Given a seed set $S$, computing $\delta_U(S)$ is \#P-hard under the IC and the LT model.
\end{theorem}
\begin{proof}
We prove by reducing from the influence spread computation problem. The reduction is the same as the one in the proof of Theorem~\ref{th:upm}. We already showed $I^G(S)=\delta_U^{G'}(S)$ for any seed set $S \subseteq V$. Therefore, the theorem follows immediately.
\end{proof}

Since computing the activity, the lower bound and the upper bound is \#P-hard, we will discuss how to estimate them in the next section.
\section{A Polling Based Method}\label{sec:alg}
Recently, a polling based algorithmic framework~\cite{borgs2014maximizing,tang2015influence} was proposed for the influence maximization problem. The framework includes two steps. In the first step, it estimates the influence spread through sampling. In the second step, it finds an approximation solution for maximizing the estimate. If we can bound the estimation error, then the solution also enjoys an approximation guarantee for the influence maximization problem. To solve the activity maximization problem, we also design a polling based method.

\subsection{Estimation}\label{sec:est}

In a social network $G$, given an information diffusion model, either the IC model or the LT model, and a seed set $S$, let $g$ be a ``live-edge'' graph instance of $G$ and $R_g(S)$ be the set of nodes reachable from $S$ in $g$.  Denote by $R_{g^T}(v)$ the reverse reachable (RR) set~\cite{tang2014influence} for node $v$ in $g$, where $g^T$ is the transpose graph~\cite{borgs2014maximizing} of $g$: $(u,v)\in g\text{ iff }(v,u)\in g^T$. We write $(u,v) \sim E$ to indicate that we randomly pick $(u,v)$ from $E$ as a sample according to a certain distribution. The meaning of $v \sim V$ is similar.

To estimate the activity, we first have the following result.
\begin{theorem}
\label{th:e1}
For any seed set $S \subseteq V$,  
$$\delta_A(S)=T\cdot {\mathop{Pr}\limits_{
                                        g\sim G,(u,v)\sim E}}\Bigg[S\cap R_{g^T}(u)\neq \emptyset \land  S\cap R_{g^T}(v)\neq \emptyset\Bigg],$$ where $T=\sum_{(u,v) \in E}{A_{u,v}}$.
\end{theorem}
\begin{proof}
    \begin{align}
    \delta_A(S)&=\mathbb{E}\Bigg[\sum\limits_{ (u,v)\in E_S}{A_{u,v}}\Bigg] \nonumber \\
               &={\sum\limits_{ (u,v)\in E}}Pr\Bigg[(u,v)\in E_S\Bigg]A_{u,v} \nonumber \\
               &={\sum\limits_{(u,v)\in E}}\mathop{Pr}\limits_{\scriptscriptstyle  g\sim G}\Bigg[u\in R_g(S) \land v\in R_g(S)\Bigg]A_{u,v} \nonumber \\
               &={\sum\limits_{(u,v)\in E}}\mathop{Pr}\limits_{\scriptscriptstyle   g\sim G}\Bigg[\exists w_1,w_2 \in S, w_1\in R_{g^T}(u) \land \nonumber \\
               &\indent \indent \indent  w_2\in R_{g^T}(v)\Bigg]A_{u,v} \nonumber \\
               &=T\cdot {\sum\limits_{(u,v)\in E}}\mathop{Pr}\limits_{\scriptscriptstyle   g\sim G}\Bigg[\exists w_1,w_2 \in S, w_1\in R_{g^T}(u) \land \nonumber \\
               &\indent \indent \indent w_2\in R_{g^T}(v)\Bigg]\frac{A_{u,v}}{T}  \label{eq:eq5}\\
               &=T\cdot {\mathop{Pr}\limits_{
                                        g\sim G,(u,v)\sim E}}\Big[\exists w_1,w_2 \in S, w_1\in R_{g^T}(u) \land \nonumber\\
                                        &\indent\indent\indent w_2\in R_{g^T}(v)\Big] \nonumber \\
               &=T\cdot{\mathop{Pr}\limits_{
                                        g\sim G,(u,v)\sim E}}\Big[S\cap R_{g^T}(u)\neq \emptyset \land S\cap R_{g^T}(v)\neq \emptyset\Big] \nonumber
    \end{align}

Eq.~\ref{eq:eq5} is the expected probability with respect to the activity distribution of edges, where the probability for edge $(u,v)$ is $\frac{A_{u,v}}{T}$.
\end{proof}

Theorem~\ref{th:e1} implies that we can estimate $\delta_A(S)$ by estimating the probability of the event $S\cap R_{g^T}(u)\neq \emptyset \land S\cap R_{g^T}(v)\neq \emptyset$. To achieve the estimation, we conduct a poll as follows. We select an edge $(u,v)$ with probability $\frac{A_{u,v}}{T}$, and run Monte Carlo simulation of the ``live-edge'' process. During the process, we record all the nodes that can reach $u$ and $v$ through ``live'' edges. Algorithm~\ref{alg:sample} summarizes the process.

One critical observation is that we do not need to conduct the ``live-edge'' process on the entire graph. Instead, we can simulate the process starting from $u$ and $v$, respectively. We only need to make sure that each edge is marked consistently as the same status (``live'' or ``blocked'') in these two simulations. We call the pair of two RR sets obtained from a poll a hyperedge.  All the generated hyperedges constitute a hypergraph $\mathcal{H}$.

\begin{figure}[t]
    \centering
\includegraphics[width=0.4\textwidth]{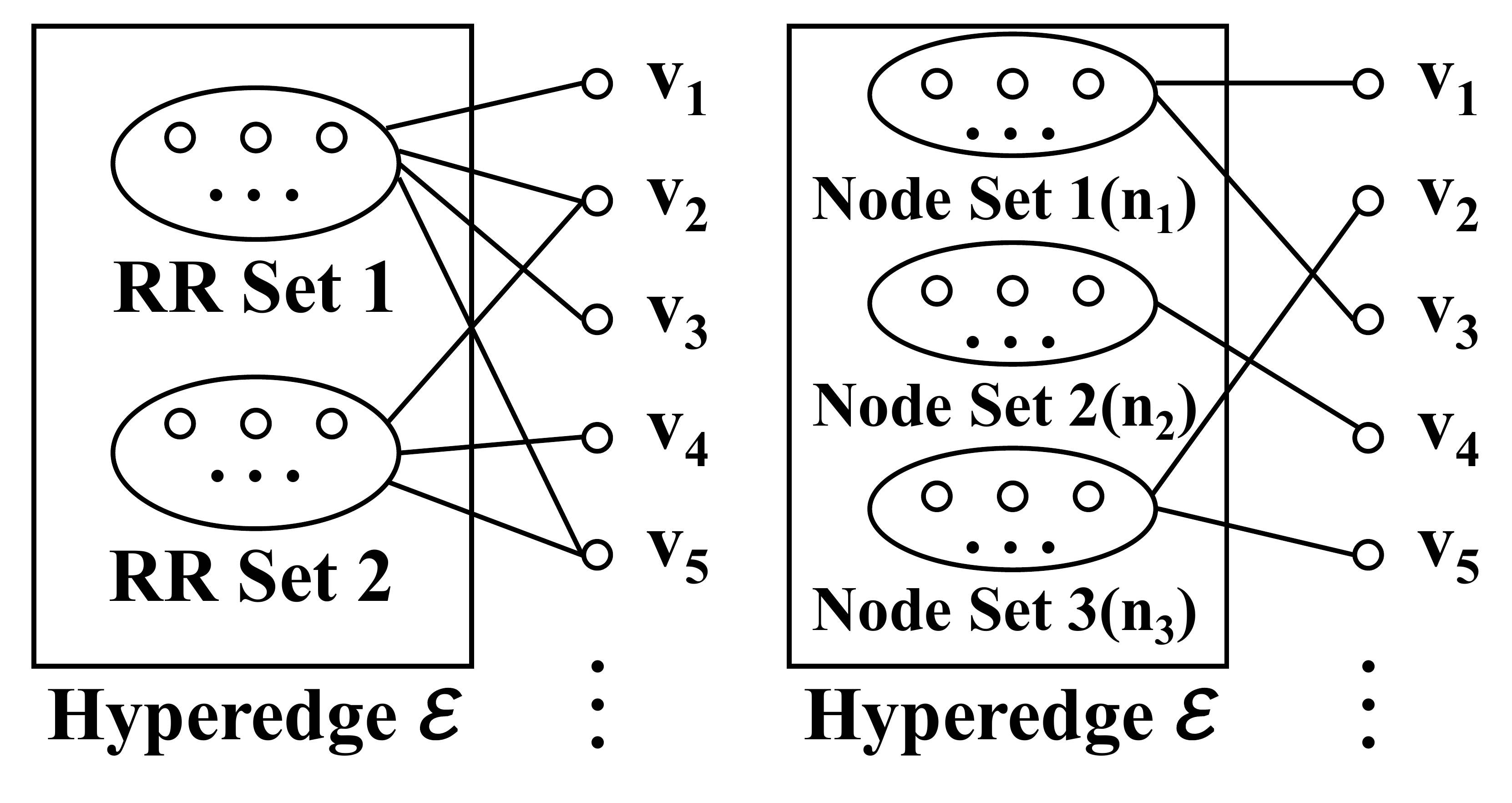}
    \caption{Hyperedge for activities}
    \label{fig:he}
\end{figure}

Denote by $m_H$ the number of the hyperedges in $\mathcal{H}$. If a node $v$ appears in both RR sets of a hyperedge $\mathcal{E}$, $\mathcal{E}$ said is to be \emph{fully covered} by $v$. If a node $v$ only appears in one of the two RR sets of a hyperedge $\mathcal{E}$, $\mathcal{E}$ is said to be \emph{partially covered} by $v$. Denote by $\mathcal{D}(S)$ the degree of the set of nodes $S$, which is the number of hyperedges in $\mathcal{H}$ that can be fully covered by $S$. According to Theorem~\ref{th:e1}, $T\cdot\frac{\mathcal{D}(S)}{m_H}$ is an unbiased estimator of $\delta_A(S)$ for any fixed $m_H$. Please note that there also exists ``combination effect'' between nodes in this case. For example, in the left part of Figure~\ref{fig:he}, $v_1$ only appears in the first RR set of hyperedge $\mathcal{E}$ and $v_4$ only appears in the second RR set. $v_1$ and $v_4$, respectively, partially covers $\mathcal{E}$. But $\mathcal{E}$ is fully covered by the combination of $v_1$ and $v_4$. Thus, similar to $\delta_A(\cdot)$, $\mathcal{D}(\cdot)$ is not submodular neither.

\begin{algorithm}[t]
\caption{Generate Hyperedges}
\label{alg:sample}
\KwIn{Social network $G=(V,E,B)$, $A$ and diffusion model $\mathcal{M}$}
\KwOut{A hyperedge $\mathcal{E}$}
\begin{algorithmic}[1]
\STATE Initialize $\mathcal{E}=(\emptyset,\emptyset)$
\STATE Pick an edge $(u,v)$ with probability $\frac{A_{u,v}}{T}$.
\STATE Generate a ``live-edge'' graph instance $g$ according to the diffusion model $\mathcal{M}$
\STATE Let $N_1=R_{g^T}(u)$ and $N_2=R_{g^T}(v)$
\STATE Let $\mathcal{E}=(N_1,N_2)$
\RETURN $\mathcal{E}$
\end{algorithmic}
\end{algorithm}

Similarly, for the lower bound and the upper bound, we have the following two results.
\begin{theorem}
\label{th:e2}
For any seed set $S \subseteq V$, $$\delta_L(S)=T\cdot {\mathop{Pr}\limits_{
                                        g\sim G,(u,v)\sim E}}\Big[S\cap (R_{g^T}(u)\cap R_{g^T}(v))\neq \emptyset\Big],$$ where $T=\sum_{(u,v) \in E}{A_{u,v}}$.
\end{theorem}
\begin{proof}
The lower bound only considers the edges whose two endpoints can be activated by the same seed node. Thus, to prove the theorem, we only need to let $w_1=w_2$ in the proof of Theorem~\ref{th:e1}, that is
\begin{equation*}
    \begin{aligned}
    \delta_L(S)=&T\cdot {\mathop{Pr}\limits_{
                                        g\sim G,(u,v)\sim E}}\Big[\exists w \in S, w\in R_{g^T}(u) \land w\in R_{g^T}(v)\Big] \\
               =&T\cdot {\mathop{Pr}\limits_{
                                        g\sim G,(u,v)\sim E}}\Big[S\cap (R_{g^T}(u)\cap R_{g^T}(v))\neq \emptyset\Big]
    \end{aligned}
\end{equation*}
\end{proof}

Using Theorem~\ref{th:e2}, we can estimate the lower bound using  essentially the same sampling process as the activity. The only difference is that there is only one node set in the hyperedge for the lower bound, that is $N_1\cap N_2$. In this case, a hyperedge $\mathcal{E}$ is covered by node $v$ if and only if $v\in N_1\cap N_2$.
\begin{theorem}
For any seed set $S \subseteq V$, $$\delta_U(S)=W\cdot{\mathop{Pr}\limits_{
                                        g\sim G,v \sim V}}\Big[S\cap R_{g^T}(v)\neq \emptyset\Big],$$ where $W=\sum_{v \in V}{w(v)}$.
\end{theorem}
\begin{proof}
The upper bound is essentially a weighted variation of the influence spread. Thus, we can apply the proof proposed in~\cite{nguyen2016cost}.
\end{proof}

There is also only one node set in the hyperedge for the upper bound. We can generate the hyperedge using the sampling method proposed in~\cite{nguyen2016cost}.

Since we can estimate the objective function ($\delta_A(\cdot)$, $\delta_L(\cdot)$ or $\delta_U(\cdot)$) by the degrees of the set of nodes, we can regard $\mathcal{H}$ as encoding an approximation to the objective function. With the estimate of the objective function, we go to the second step of the polling based framework, that is, maximizing the estimate. To achieve this goal, we adopt the simple but powerful greedy strategy, which picks the node with the largest marginal gain (the increase of degree in $\mathcal{H}$ for our case) iteratively.  Next, we show how to efficiently implement a greedy strategy on the hypergraph.

\subsection{Efficient Implementation of the Greedy Strategy}

For the lower bound and the upper bound, there is only one node set in each hyperedge. Thus, we can use the standard greedy algorithm for maximum coverage problem to obtain an approximate solution~\cite{tang2014influence}. However, there are two node sets in the hyperedge for the activity maximization problem. A hyperedge can be fully or partially covered by a node or a node set. Thus, we cannot directly apply the greedy strategy. To tackle this issue, here we discuss how to efficiently implement the greedy strategy on the hypergraph.

First, we store the original hyperedges of two RR sets in a more efficient manner. There are three sets, $n_1$, $n_2$ and $n_3$ for each hyperedge $\mathcal{E}$, where $n_1$ and $n_2$ are the sets of nodes that can only cover the first and second RR set of $\mathcal{E}$, respectively, and $n_3$ is the set of nodes that can cover both two RR sets of $\mathcal{E}$. Fig.~\ref{fig:he} illustrates the idea.

Then, we build an inverted index for each node. There are three sets, $e_1$, $e_2$ and $e_3$ for each node $v$, where $e_1$ and $e_2$ are the sets of hyperedges whose first and second RR set can be covered by $v$, respectively, and $e_3$ is the set of hyperedges that can be fully covered by $v$.

\begin{figure}[t]
    \centering
\includegraphics[width=0.4\textwidth]{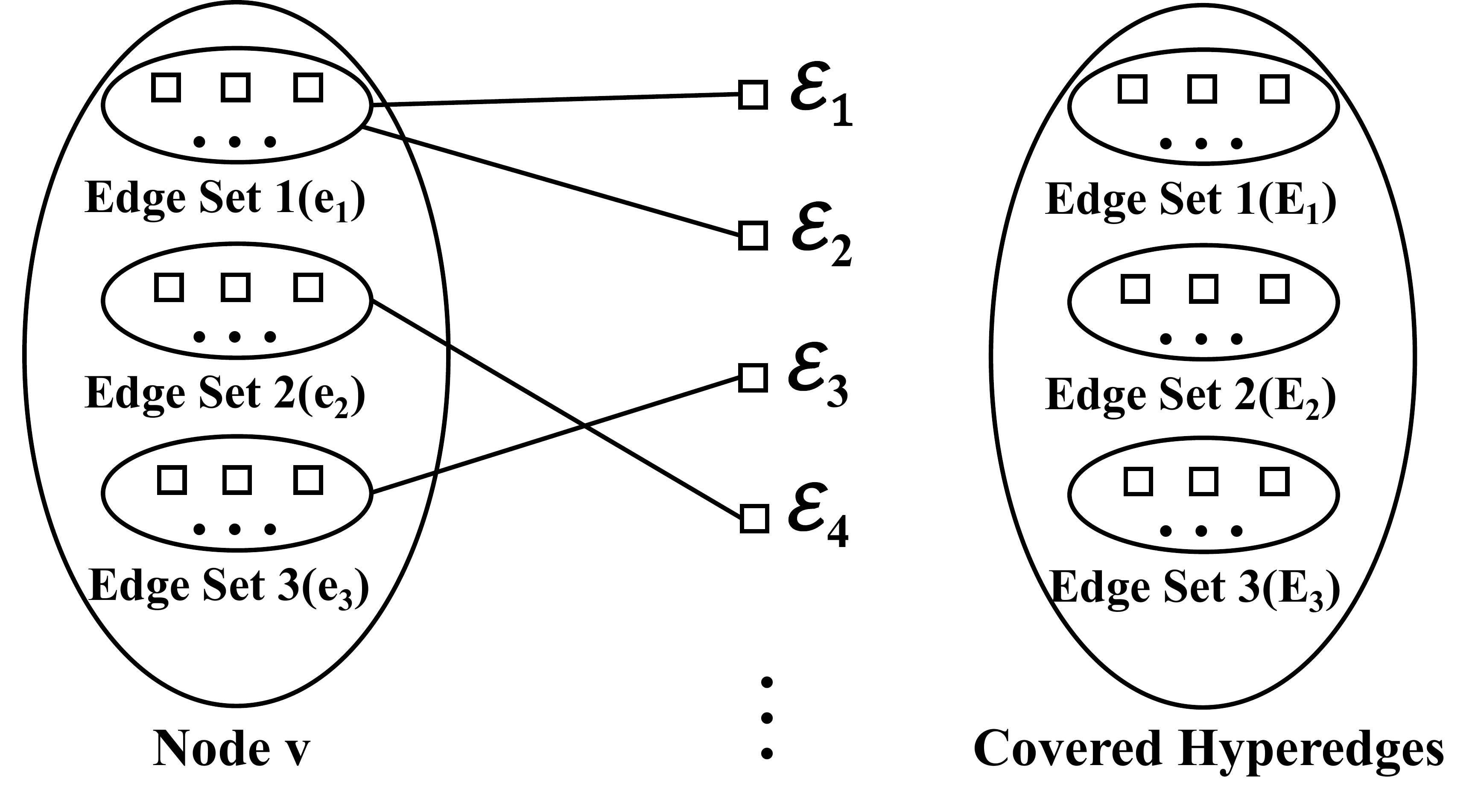}
    \caption{Data Structures}
    \label{fig:ds}
\end{figure}

Third, we maintain a global data structure to record the current covered hyperedges. There are also three sets, $E_1$, $E_2$ and $E_3$, in this data structure, where $E_1$ and $E_2$ are the sets of hyperedges whose first and second RR sets have been covered, respectively, and $E_3$ is the set of hyperedges that have been fully covered. Fig.~\ref{fig:ds} shows these two data structures. With these data structures, we have the following fact.

\begin{fact}
\label{fa:mg}
Given a seed set $S$, for each vertex $v \in V\setminus S$, the marginal gain $\mathcal{D}(S\cup \{v\})-\mathcal{D}(S)$ is
\begin{equation}
\label{eq:mg_def}
MG(v)=|v.e_3\setminus E_3|+|v.e_1\cap E_2|+|v.e_2\cap E_1|
\end{equation}

\noindent{\sc Rationale.} \em
If we add a node $v$ to the current seed set $S$, the newly covered hyperedges can be divided into two groups. The first group is the hyperedges that can be covered by $v$ alone but not covered by $S$, that is $v.e_3\setminus E_3$. The second group is the hyperedges that are partially covered by $S$ and are fully covered if $v$ is added to $S$, that is $v.e_1\cap E_2$ and $v.e_2\cap E_1$.
\boxend
\end{fact}

Fact~\ref{fa:mg} implies that we can pick the node with the largest marginal gain in each iteration and then incrementally update the marginal gains of the rest nodes. Algorithm~\ref{alg:max_cover} describes the details.

Here, we briefly explain how to incrementally update the marginal gain. Assuming $E_1$, $E_2$ and $E_3$ are updated to $E'_1$, $E'_2$, and $E'_3$, respectively, we update the marginal gains as follows. For each hyperedge $\mathcal{E} \in E'_1\setminus E_1$, we increase the marginal gains of the nodes in $\mathcal{E}.n_2$ by 1. For each hyperedge $\mathcal{E} \in E'_2\setminus E_2$, we increase the marginal gains of the nodes in $\mathcal{E}.n_1$ by 1. For each hyperedge $\mathcal{E} \in E'_3\setminus E_3$, we first decrease the marginal gains of the nodes in $\mathcal{E}.n_3$ by 1. Then, we decrease the marginal gains of the nodes in $\mathcal{E}.n_2$ by 1 if $\mathcal{E} \in E_1$, and decrease the marginal gains of the nodes in $\mathcal{E}.n_1$ by 1 if $\mathcal{E} \in E_2$.

Now, the only remaining question is to decide how many hyperedges we need to sample, which will be addressed next.

\begin{algorithm}[t]
\caption{Maximum Coverage on Hypergraph}
\label{alg:max_cover}
\KwIn{Social network $G$, Hypergraph $\mathcal{H}$ and budget $k$}
\KwOut{Seed set $S$}
\begin{algorithmic}[1]

\STATE Initialize $S=E_1=E_2=E_3=\emptyset$
\FOR{ $v \in V$ }
\STATE $MG(v)=|v.e|$
\ENDFOR
\WHILE{$|S|<k$}
\STATE $v=\argmax_{u\in V\setminus S}{MG(u)}$
\STATE $S=S\cup\{v\}$
\STATE update $E_1,E_2,\text{ and }E_3$
\FOR{ $u\in V\setminus S$ }
\STATE update $MG(u)$ according to Eq.~\ref{eq:mg_def}
\ENDFOR
\ENDWHILE
\RETURN $S$
\end{algorithmic}
\end{algorithm}

\subsection{Sample Complexity} \label{sec:sample}

In this subsection, we discuss how to use a sample of proper size to restrict the estimate error of the activity, the lower bound and the upper bound. With the technique, we show that the polling algorithm can provide an approximate solution to maximizing the lower bound and the upper bound.

To bound the estimate error of the polling method, we have the following lemma from~\cite{dagum2000optimal}.

\begin{lemma}
\label{le:mc}
Let $Z_1,Z_2, \ldots$ be independently and identically distributed according to Z in the interval $[0,1]$ with mean $\mu_Z$. Let $S=\sum_{i=0}^{N}{Z_N}$ and $\hat{\mu}_Z=\frac{S}{N}$. Let $\Upsilon=4(e-2)\frac{\ln(2/\delta)}{\epsilon^2}$ and $\Upsilon_1=1+(1+\epsilon)\Upsilon$. If $N$ is the number of samples when $S>=\Upsilon_1$, then
$Pr[|\hat{\mu}_Z-\mu_Z|\leq \epsilon\mu_Z]>1-\delta$ and
$\mathbb{E}[N]\leq \Upsilon_1/\mu_Z $.
\end{lemma}
Lemma~\ref{le:mc} provides a stopping condition for the sampling process. Given a seed set $S$, we can keep sampling hyperedges until $\mathcal{D}(S)\geq \Upsilon_1$. Then, $T\cdot\frac{\mathcal{D}(S)}{m_H}$ is an $(\epsilon,\delta)$ estimation~\cite{mitzenmacher2005probability} of $\delta_A(S)$. The analysis is similar in the cases of the lower bound and the upper bound.

Nguyen \textit{et~al.}~\cite{nguyen2016stop} analyzed the conditions that the polling algorithmic framework must meet to obtain an approximation solution. Let $S^\ast$ be the optimal seed set and $\hat{S}$ be the seed set returned by the greedy strategy on the estimate of the objective function $f(\cdot)$ ($\delta_L(\cdot)$ or $\delta_U(\cdot)$). Denote by $\hat{f}(\cdot)$ the estimate of the objective function $f(\cdot)$. The conditions are
\begin{equation}
\label{eq:c1}
Pr[\hat{f}(\hat{S})\leq (1+\epsilon_1)f(\hat{S})] \geq 1-\delta_1
\end{equation}
\begin{equation}
\label{eq:c2}
Pr[\hat{f}(S^\ast)\geq (1-\epsilon_2)f(S^\ast)] \geq 1-\delta_2
\end{equation}
where $\delta_1+\delta_2\leq \delta$ and $\epsilon_1+(1-1/e)\epsilon_2\leq \epsilon$. Let $N$ be the number of samples such that both Eq.~\ref{eq:c1} and Eq.~\ref{eq:c2} are guaranteed. Then we have the following lemma from~\cite{nguyen2016stop}.
\begin{lemma}
\label{le:ssa}
Given a social network $G$, if the number of hyperedges $m_H\geq N$, then the polling algorithm returns $\hat{S}$ satisfying $Pr[f(\hat{S})\geq (1-1/e-\epsilon)f(S^\ast)] \geq 1-\delta$ and $\hat{S}$ is an $(1-1/e-\epsilon)$ approximate solution.
\end{lemma}

Using Lemmas~\ref{le:mc} and~\ref{le:ssa}, to obtain an approximation solution to maximizing the lower bound or upper bound, we can keep sampling hyperedges and checking if the conditions are met. Algorithm~\ref{alg:ssa} from~\cite{nguyen2016stop} describes the process. Please note that, in Lines~4 and~8 of Algorithm~\ref{alg:ssa}, we adopt the standard greedy algorithm for the maximum coverage problem to get $\hat{S}$.

\begin{algorithm}[t]
\caption{SSA Algorithm}
\label{alg:ssa}
\begin{algorithmic}[1]

\STATE Compute $\Upsilon_1$ according to $\epsilon_1,\epsilon_2,\delta_1,\delta_2$
\STATE $m_H=\Upsilon_1$
\STATE Generate $m_H$ hyperedges and add to $\mathcal{H}$
\STATE $\hat{S}=$maximum coverage on $\mathcal{H}$
\WHILE{$\mathcal{D}(\hat{S}) <\Upsilon_1 $}
\STATE Generate $m_H$ hyperedges and add to $\mathcal{H}$
\STATE $m_H=2m_H$
\STATE $\hat{S}=$maximum coverage on $\mathcal{H}$
\ENDWHILE
\RETURN $\hat{S}$
\end{algorithmic}
\end{algorithm}

Using Algorithm~\ref{alg:ssa}, we can provide a $(1-1/e-\epsilon)$ approximation solution to maximizing the lower bound and the upper bound with probability of at least $1-\delta$. But we must point out that the analysis does not hold for the activity maximization problem. This is because a necessary condition of the polling algorithmic framework is that we can approximate the estimate using the greedy strategy. The condition is not met in the case of the activity maximization problem, since the estimate of the activity is not submodular. Thus, the polling algorithm cannot provide an approximation solution to the activity maximization problem. But it is still a good heuristic for the activity maximization problem. Furthermore, by combining the approximation algorithm for the lower bound and the upper bound, we can derive a data dependent approximation scheme for the activity maximization problem.

\subsection{Data Dependent Approximation}
There is no general way to optimize or approximate a non-submodular function. Lu \textit{et~al.}~\cite{lu2015competition} proposed a sandwich approximation strategy, which approximates the objective function by approximating its lower bound and upper bound. The sandwich approximation strategy works as follows. First, we find a solution to the original problem with any strategy. Second, we find an approximate solution to the lower bound and the upper bound, respectively. Last, we return the solution that has the best result for the original problem.

Here, we extend the strategy to the case in which the objective function is intractable and have the following result.

\begin{algorithm}[t]
\caption{Sandwich Approximation Framework}
\label{alg:sandwich}
\begin{algorithmic}[1]

\STATE Let $S_U$ be a $\alpha$ approximation to the upper bound
\STATE Let $S_L$ be a $\beta$ approximation to the lower bound
\STATE Let $S_A$ be a solution to the original problem
\STATE $\hat{\delta}_A(\cdot)$ is a multiplicative $\gamma$-error estimate of $\delta_A(\cdot)$
\STATE $S=\argmax_{S_0\in \{S_U,S_L,S_A\}}{\hat{\delta}_A(S_0)}$
\RETURN $S$
\end{algorithmic}
\end{algorithm}
\begin{theorem}
\label{th:sandwich}
Let $S$ be the seed set returned by Algorithm~\ref{alg:sandwich}, then we have
\begin{equation}
\delta_A(S)\geq \max\Big\{\frac{\delta_A(S_U)}{\delta_U(S_U)}\alpha,\frac{\delta_L(S_L^\ast)}{\delta_A(S_A^\ast)}\beta\Big\}\frac{1-\gamma}{1+\gamma}\delta_A(S_A^\ast)
\end{equation}
\end{theorem}
\begin{proof}
Let $S_L^\ast$, $S_U^\ast$ and $S_A^\ast$ be the optimal solutions to maximizing the lower bound, the upper bound and the  activity, respectively. Then, we have
 \begin{equation*}
 \begin{aligned}
  \delta_A(S_U)&=\frac{\delta_A(S_U)}{\delta_U(S_U)}\delta_U(S_U)\geq \frac{\delta_A(S_U)}{\delta_U(S_U)}\cdot\alpha\cdot\delta_U(S_U^\ast)\\
               &\geq \frac{\delta_A(S_U)}{\delta_U(S_U)}\cdot\alpha\cdot\delta_U(S_A^\ast) \\
               &\geq \frac{\delta_A(S_U)}{\delta_U(S_U)}\cdot\alpha\cdot\delta_A(S_A^\ast) \\
   \delta_A(S_L)&\geq \delta_L(S_L)\geq \beta\cdot\delta_L(S_L^\ast)\\
                &\geq \frac{\delta_L(S_L^\ast)}{\delta_A(S_A^\ast)}\cdot\beta\cdot\delta_A(S_A^\ast)
    \end{aligned}
 \end{equation*}
 Let $S_{max}=\argmax_{S_0\in \{S_U,S_L,S_A\}}{\delta_A(S_0)}$, then
 $$\delta_A(S_{max})\geq \max\Big\{\frac{\delta_A(S_U)}{\delta_U(S_U)}\alpha,\frac{\delta_L(S_L^\ast)}{\delta_A(S_A^\ast)}\beta\Big\}\delta_A(S_A^\ast)$$
 Since $\forall S_0\in \{S_U,S_L,S_A\},|\hat{\delta}_A(S_0)-\delta_A(S_0)| \leq \gamma\delta_A(S_0)$, we have
 $(1+\gamma)\delta_A(S)\geq (1-\gamma)\delta_A(S_{max})$. It follows that
 $$\delta_A(S)\geq \frac{(1-\gamma)}{(1+\gamma)}\delta_A(S_{max}) $$
\end{proof}

Theorem~\ref{th:sandwich} indicates that we can approximate the activity maximization problem within a factor that is dependent on the data. Since it is \#P-hard to compute $\delta_A(\cdot)$ and $\delta_U(\cdot)$, and is NP-hard to find $S_L^\ast$ and $S_A^\ast$, we cannot compute the exact approximation factor. But we can estimate $\frac{\delta_A(S_U)}{\delta_U(S_U)}$ by computing its lower bound $\frac{(1-\gamma)\hat{\delta}_A(S_U)}{(1+\gamma)\hat{\delta}_U(S_U)}$. It follows that $\frac{(1-\gamma)^2}{(1+\gamma)^2}\cdot\alpha\cdot\frac{\hat{\delta}_A(S_U)}{\hat{\delta}_U(S_U)}$ is a computable lower bound of the approximation factor.

Now, we put all the pieces of the puzzle together. We first adopt the polling algorithm to maximize the lower bound and the upper bound. As discussed in Section~\ref{sec:sample}, it provides $(1-1/e-\epsilon)$ approximate solutions to the lower bound and the upper bound, respectively. Consequently, we have $\alpha=\beta=(1-\frac{1}{e}-\epsilon)$ in Algorithm~\ref{alg:sandwich}. Then, we also use the polling algorithm to get a heuristic solution ($S_A$) to the activity maximization problem. Last, we get a $(\gamma,\delta)$ estimation of $\delta_A(\cdot)$ based on Lemma~\ref{le:mc} to complete Line~5 of Algorithm~\ref{alg:sandwich}. According to Theorem~\ref{th:sandwich}, the sandwich algorithm returns a seed set $S$ such that $$\delta_A(S)\geq \max\Big\{\frac{\delta_A(S_U)}{\delta_U(S_U)},\frac{\delta_L(S_L^\ast)}{\delta_A(S_A^\ast)}\Big\}\frac{1-\gamma}{1+\gamma}(1-\frac{1}{e}-\epsilon) \delta_A(S_A^\ast)$$

\section{Related Work}\label{sec:rel}
Domingos and Richardson~\cite{domingos2001mining} first exploited the influence between users in social networks for viral marketing.  Kempe \textit{et~al.}~\cite{kempe2003maximizing} formulated the problem as a discrete optimization problem, which is also well known as the influence maximization problem.
The influence maximization problem aims to optimize the influence spread (the expected number of activated nodes) in a given information diffusion model, such as the IC model and the LT model. Due to its important applications in viral marketing and some other areas, it has drawn much attention from both academia and industry~\cite{leskovec2007cost,du2013scalable,cohen2014sketch,lucier2015influence,yang2016continuous,chen2015online}.

Under the IC model and the LT model, Kempe \textit{et~al.}~\cite{kempe2003maximizing} proved that influence maximization is NP-hard.  Moreover, Chen \textit{et~al.}~\cite{chen2010scalable,chen2010scalablelt} proved that computing influence spread is \#P-hard. Thus, many heuristic algorithms were proposed to solve the problem under these two models~\cite{chen2010scalable,chen2010scalablelt,goyal2011simpath,liu2014influence,cheng2014imrank}. Recently, a polling based method~\cite{borgs2014maximizing} was proposed for influence maximization. Unlike the previous heuristic algorithms, this method can provide a solution with provable approximation guarantee. Later, Tang \textit{et~al.}~\cite{tang2014influence,tang2015influence} reduced the sample complexity and improved the efficiency. Nguyen \textit{et~al.}~\cite{nguyen2016stop} further sped up the algorithm with a different bounding technique~\cite{dagum2000optimal}. In this paper, we extend this algorithmic framework to solve our activity maximization problem in a non-trivial way.

A series of extensions to the influence maximization problem have been studied.  For example, Goyal \textit{et~al.}~\cite{goyal2011data} proposed a data based approach to influence maximization based on a credit distribution model. Instead of maximizing the influence spread under some propagation models with respect to some learned parameters, they tried to find influential nodes from the action log data directly. Chen \textit{et~al.}~\cite{chen2012time} considered the time-delay aspect of influence diffusion and studied the influence maximization with time-critical constraint. Tang \textit{et~al.}~\cite{tang2014diversified} studied the problem of maximizing the influence spread and the diversity of the influenced crowd simultaneously. Bhagat \textit{et~al.}~\cite{bhagat2012maximizing} argued that product adoption should be distinguished from influence spread in viral marketing, as influence spread is essentially used as ``proxy'' for product adoption. Wang \textit{et~al.}~\cite{wang2015maximizing} distinguished the information coverage and information propagation, and proposed a new optimization objective that includes the values of the informed nodes. All these extensions were from the perspective of nodes and tried to exploit the values of nodes as separate individuals in different diffusion models and different problem settings. They did not consider activity strengths on edges in their objectives.  In contrast, our problem captures the interactions among nodes and enables different (often orthogonal) applications of information diffusion.

\section{Experiments}\label{sec:exp}
In this section, we evaluate our algorithm via a series of experiments on three real-world data sets.

\subsection{Settings}

We ran our experiments on three real-world data sets, which are available at the SNAP website (\url{http://snap.stanford.edu}). Tab.~\ref{tab:data} shows the statistics of the data sets.

\begin{table}[t]
\centering\small
\begin{tabular}{|l|c|c|c|c|}
\hline
Network & \# Vertices & \# Edges & Average degree \\ \hline
HepPh & 12,008 & 118,521 & 9.9 \\ \hline
DBLP & 317,080 & 1,049,866 & 3.3 \\ \hline
LiveJournal & 3,997,962 & 34,681,189 & 8.7\\ \hline
\end{tabular}
\caption{The statistics of the data sets.}
\label{tab:data}
\end{table}

The propagation probability $B_{u,v}$ for the IC model and the influence weight for the LT model of an edge $(u,v)$ is set to $\frac{1}{degree(v)}$, as widely used in literatures~\cite{chen2013information}. Since activity maximization is a novel problem that has not been studied in the past, nor the activity of each edge, we do not have any real data for the purpose of experiments. Thus, we verify our algorithm with two synthetic activity settings. In the first case, we uniformly set $A_{u,v}$ to $1$ for each edge $(u,v)$. In the second case, we set $A_{u,v}$ to the value of the diffusion parameter $B_{u,v}$. The intuition is that there might be more interactions between $u$ and $v$ if $u$ is more likely to activate $v$.  For the approximation parameters, we set $\epsilon=0.1$, $\delta=0.001$ and $\gamma=0.05$ for all data sets.

We compare the proposed algorithm, referred as Sandwich, with three heuristic algorithms: InfMax, Degree and PageRank. InfMax returns the nodes for influence maximization. We followed the implementation reported in~\cite{nguyen2016stop}. Degree returns the nodes with high degrees. PageRank returns the nodes with high PageRank~\cite{page1999pagerank} scores.

We implemented our algorithm and the baselines in Java. All experiments were conducted on a PC computer with a 3.4GHZ Intel Core i7-3770 processor and 32 GB memory, running Microsoft Windows 7.

\subsection{Effectiveness}

\begin{figure*}[t]
    \centering
    \subfigure[HepPh-IC-uniform]{\includegraphics[width=43mm]{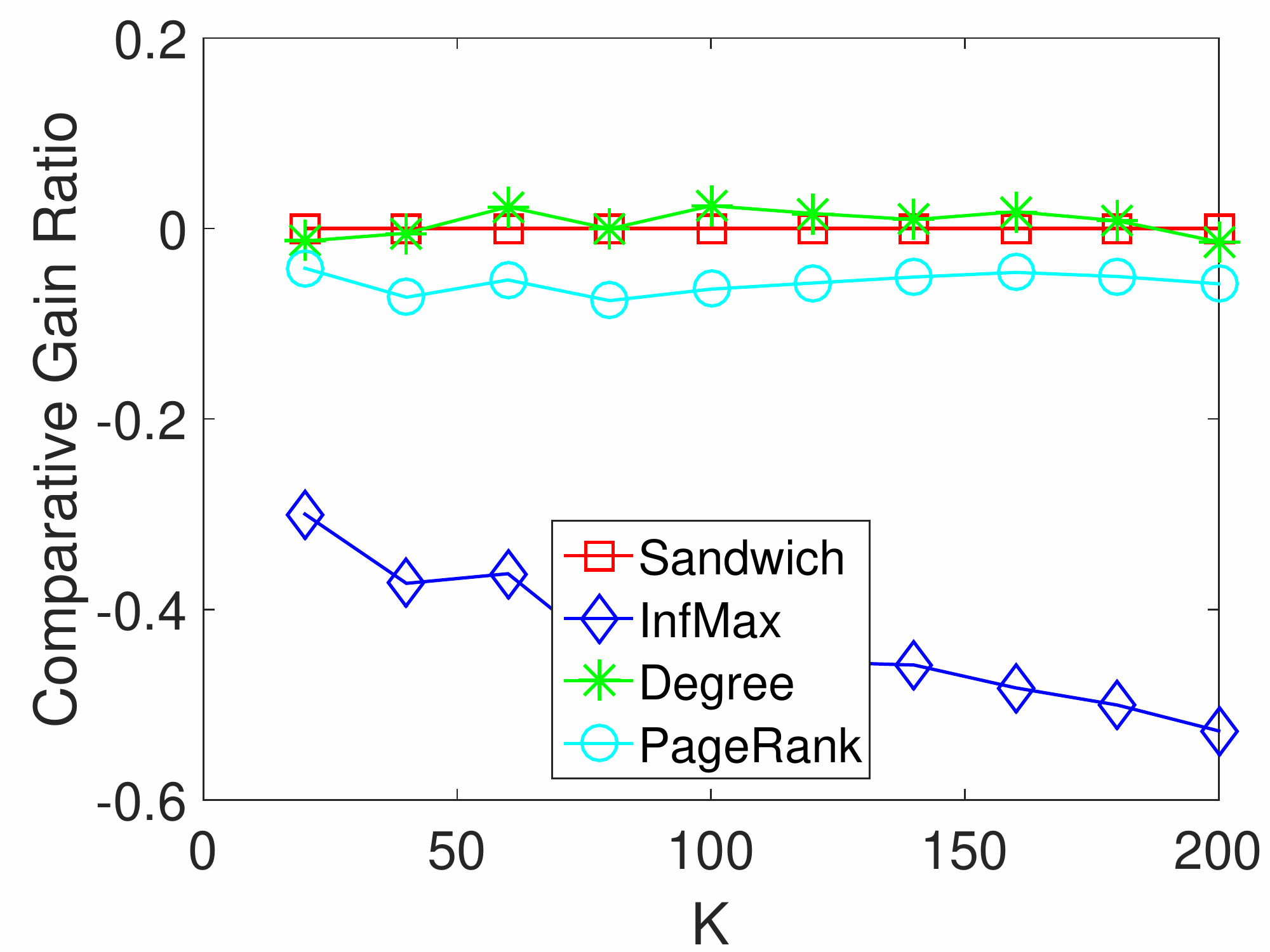}}
    \subfigure[HepPh-IC-diffusion]{\includegraphics[width=43mm]{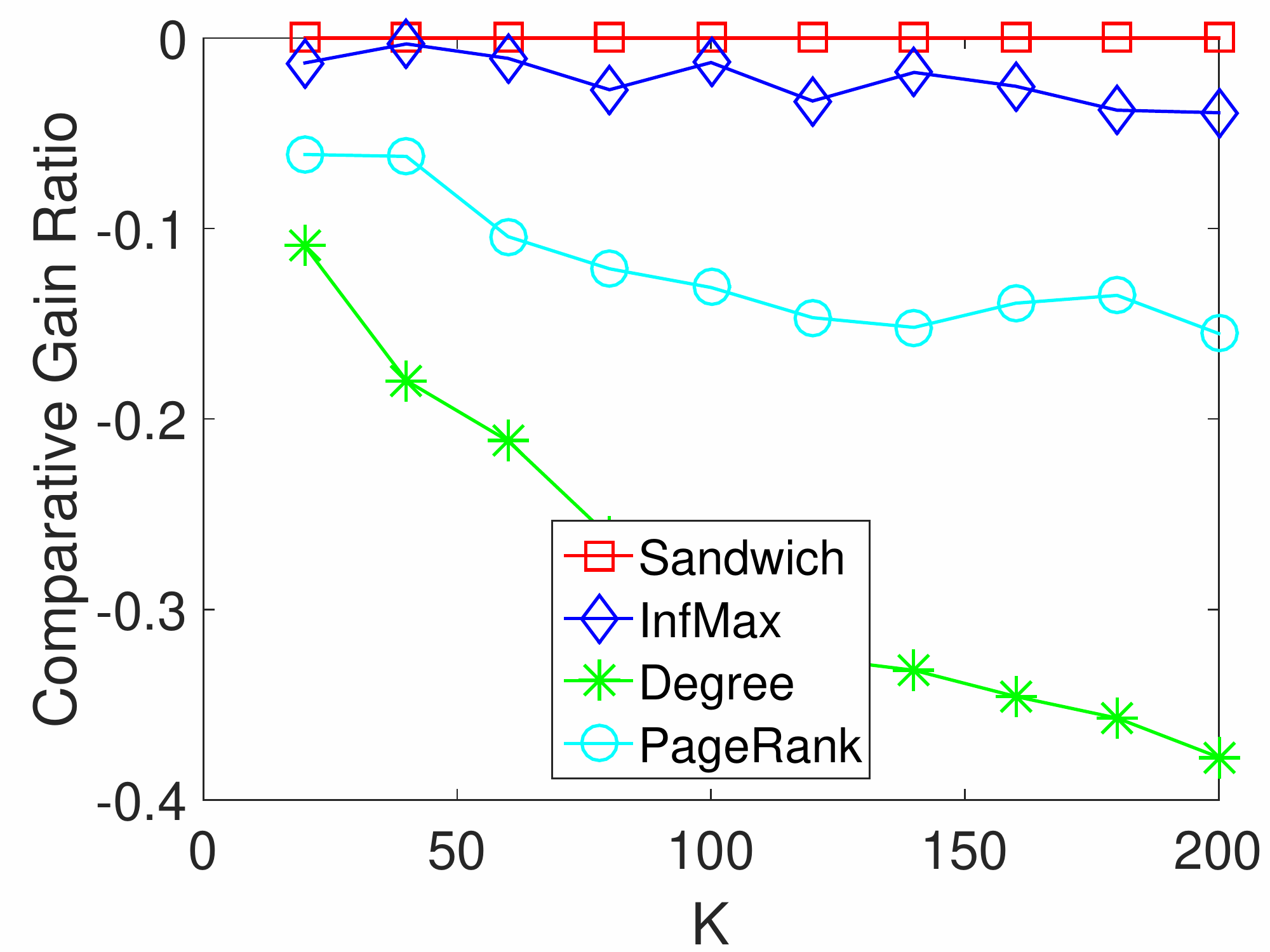}}
    \subfigure[HepPh-LT-uniform]{\includegraphics[width=43mm]{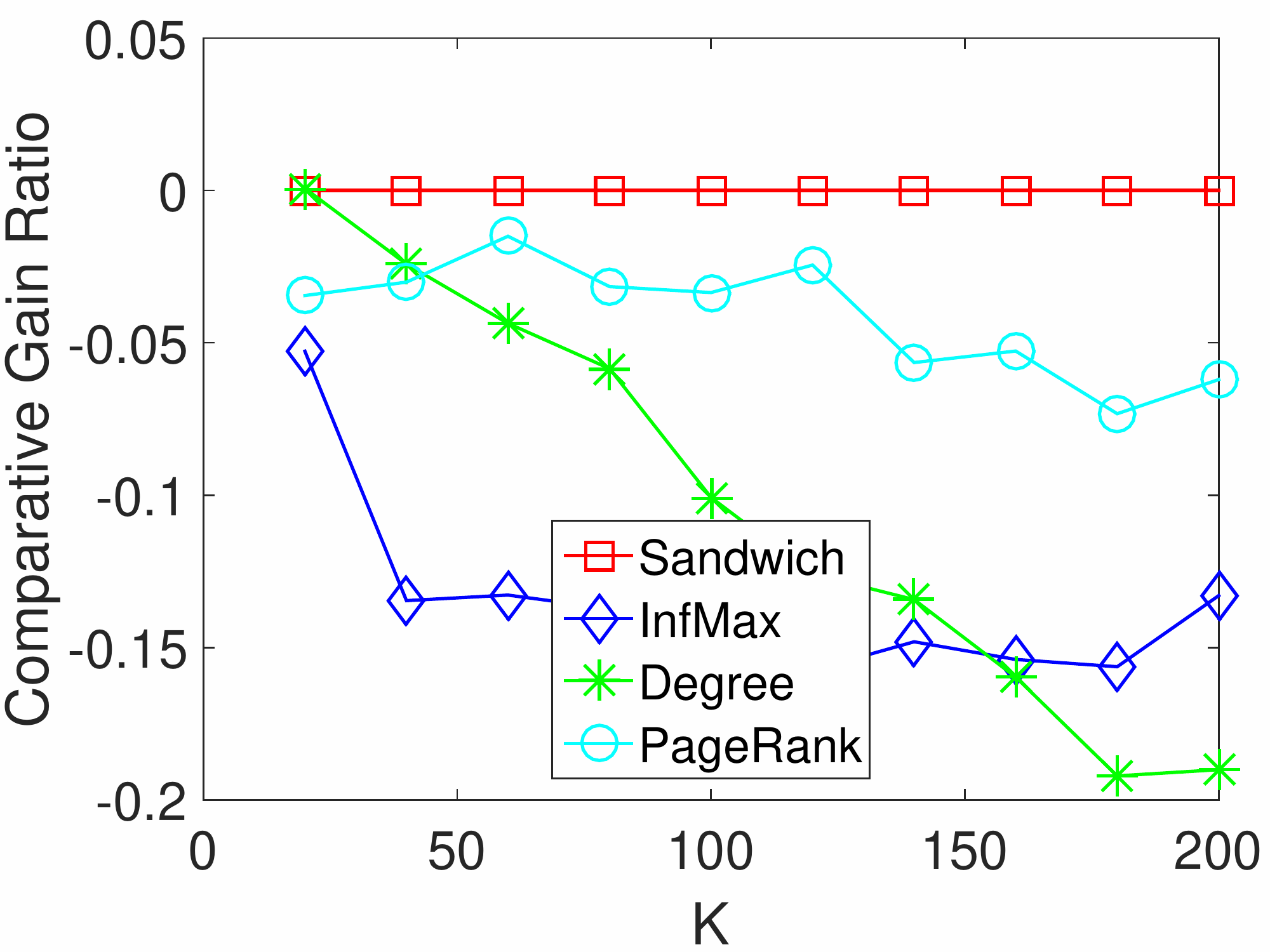}}
    \subfigure[HepPh-LT-diffusion]{\includegraphics[width=43mm]{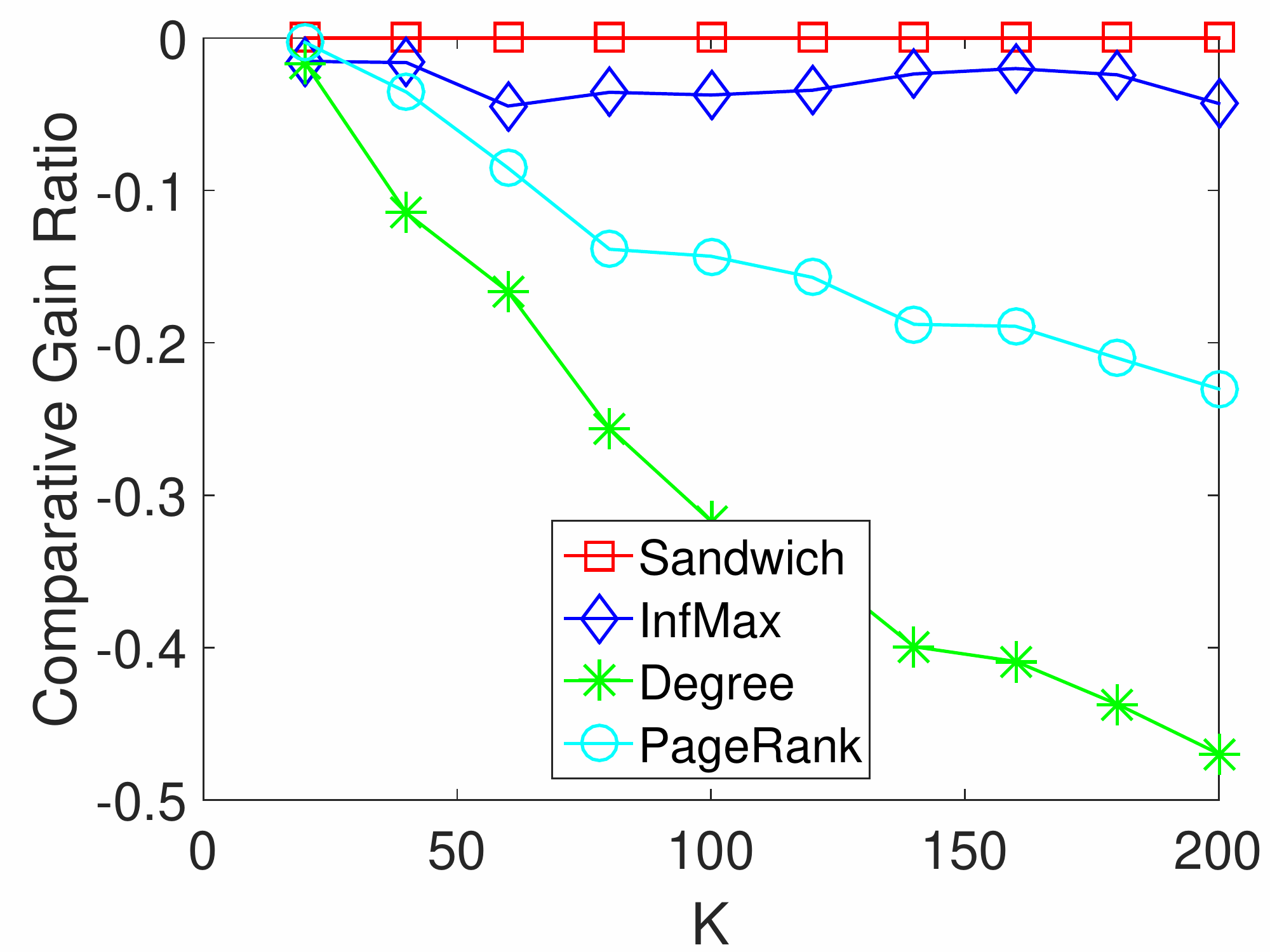}}

    \subfigure[DBLP-IC-uniform]{\includegraphics[width=43mm]{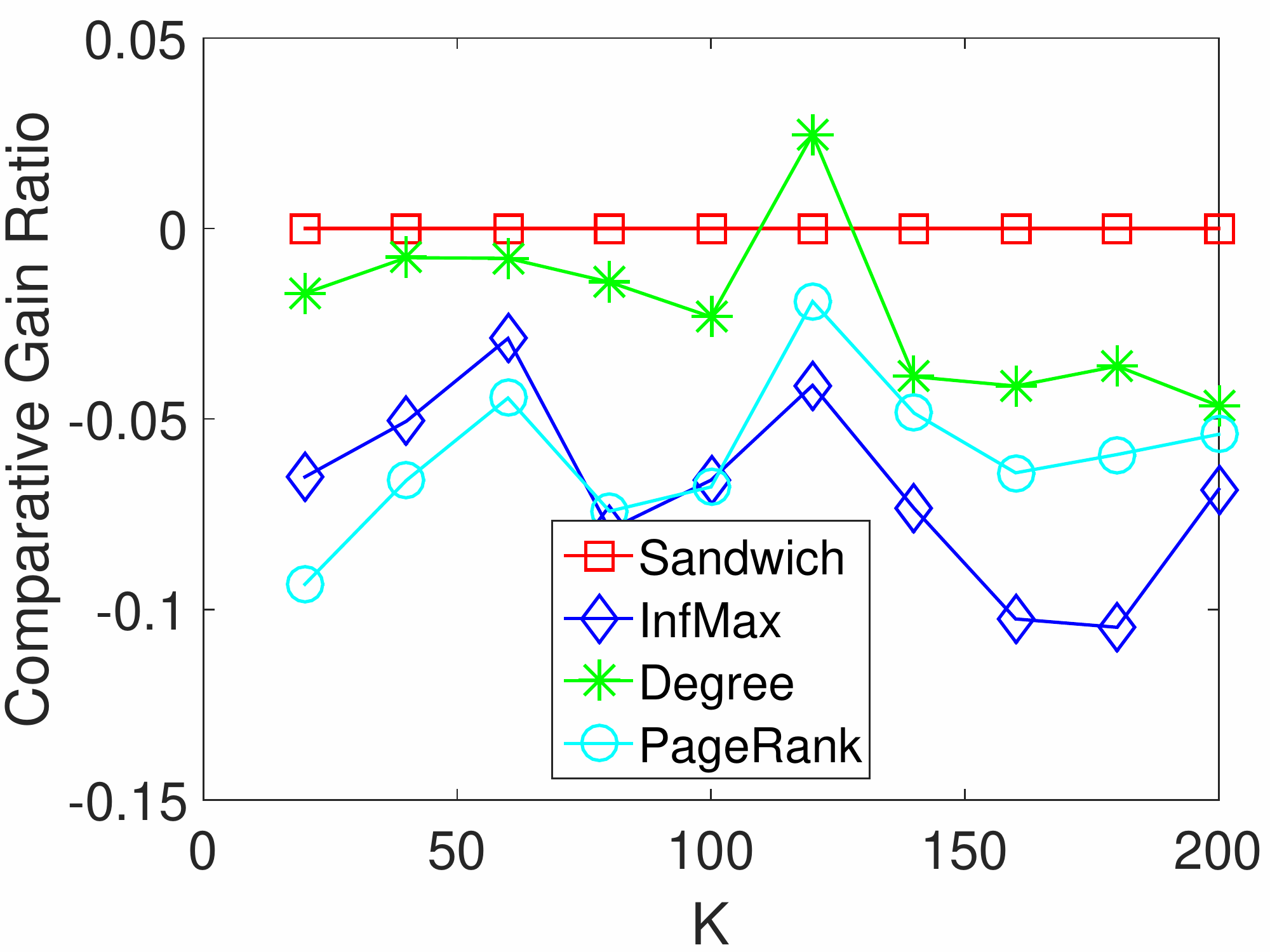}}
    \subfigure[DBLP-IC-diffusion]{\includegraphics[width=43mm]{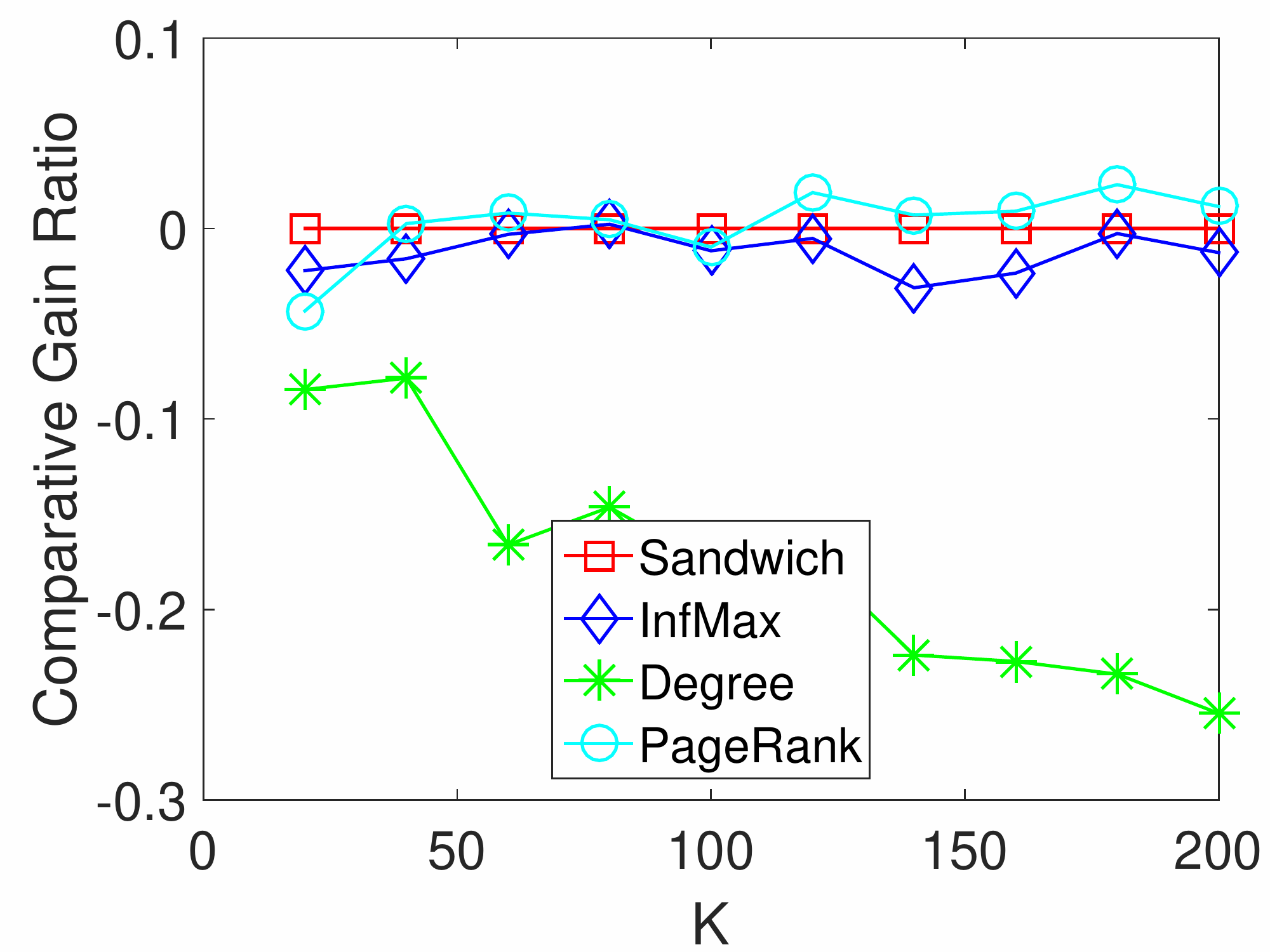}}
    \subfigure[DBLP-LT-uniform]{\includegraphics[width=43mm]{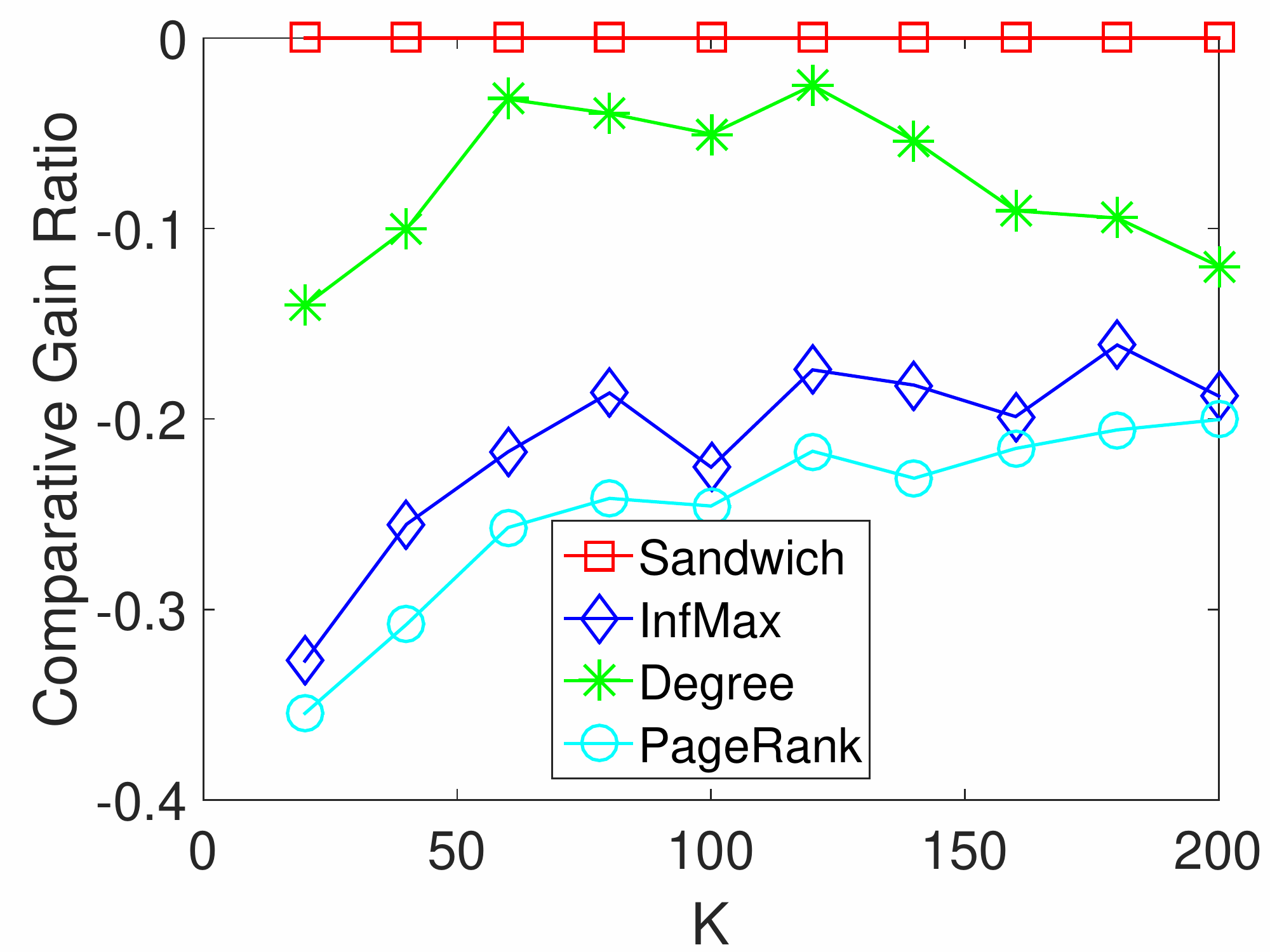}}
    \subfigure[DBLP-LT-diffusion]{\includegraphics[width=43mm]{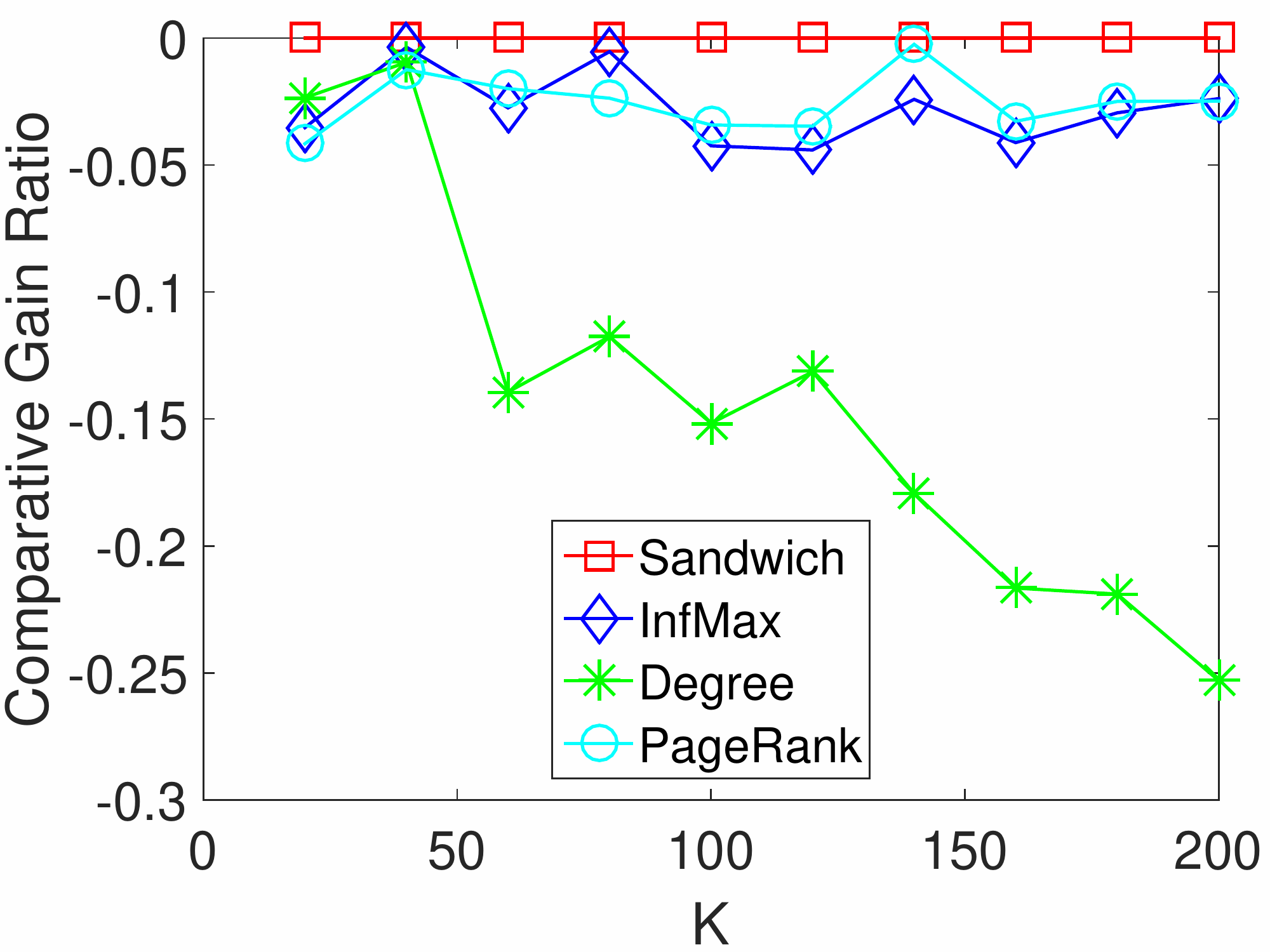}}

    \subfigure[LiveJournal-IC-uniform]{\includegraphics[width=43mm]{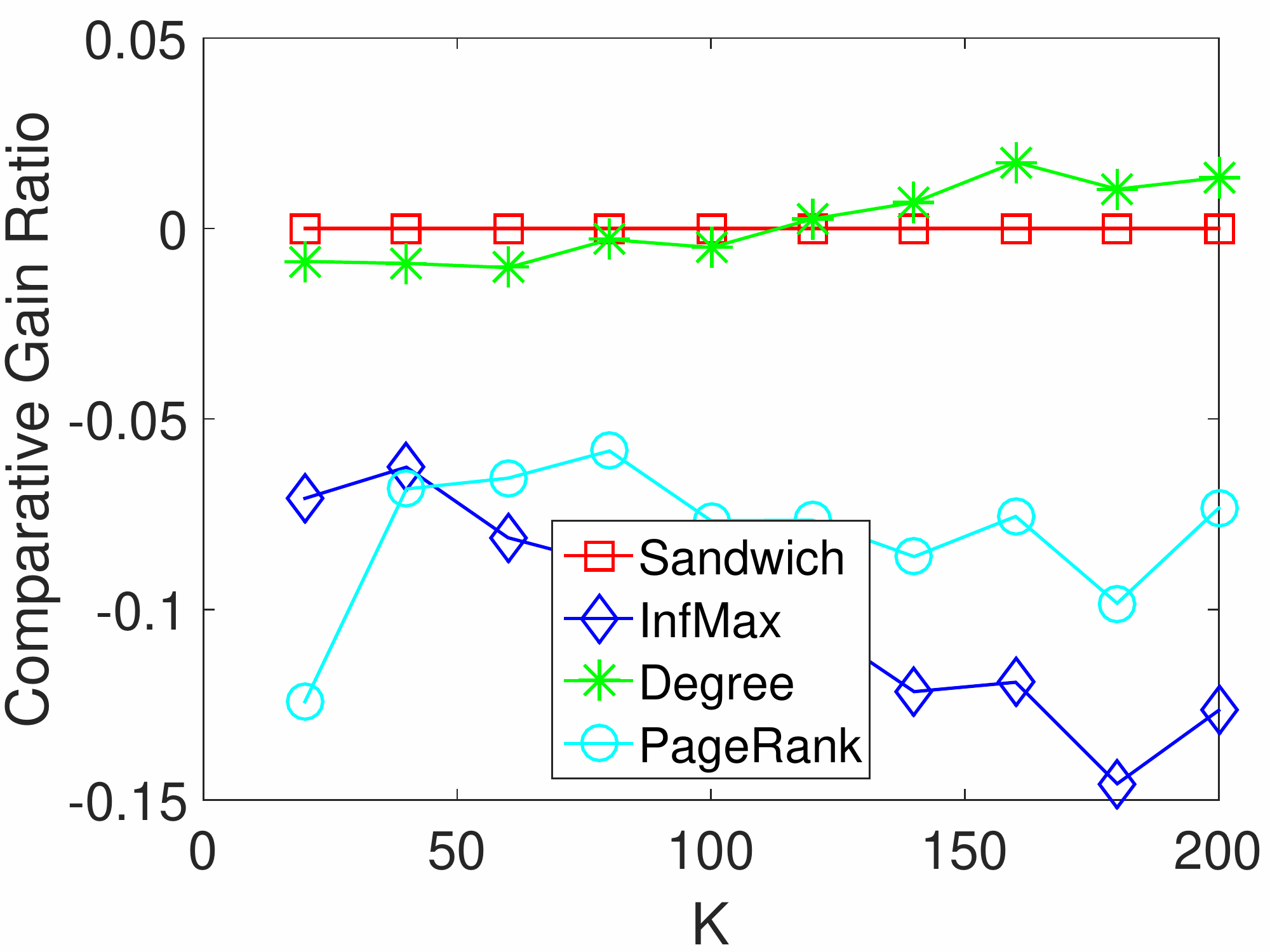}}
    \subfigure[LiveJournal-IC-diffusion]{\includegraphics[width=43mm]{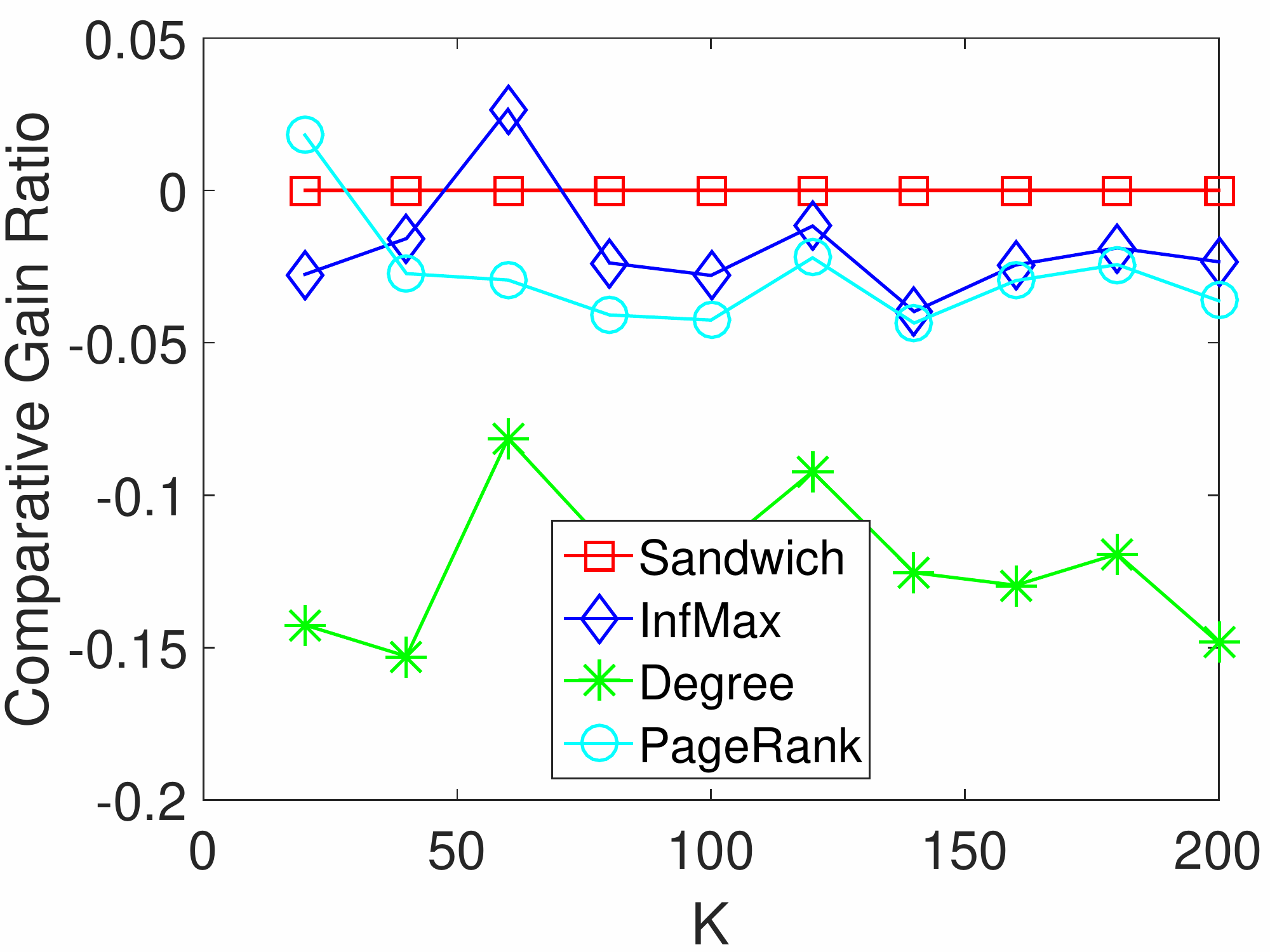}}
    \subfigure[LiveJournal-LT-uniform]{\includegraphics[width=43mm]{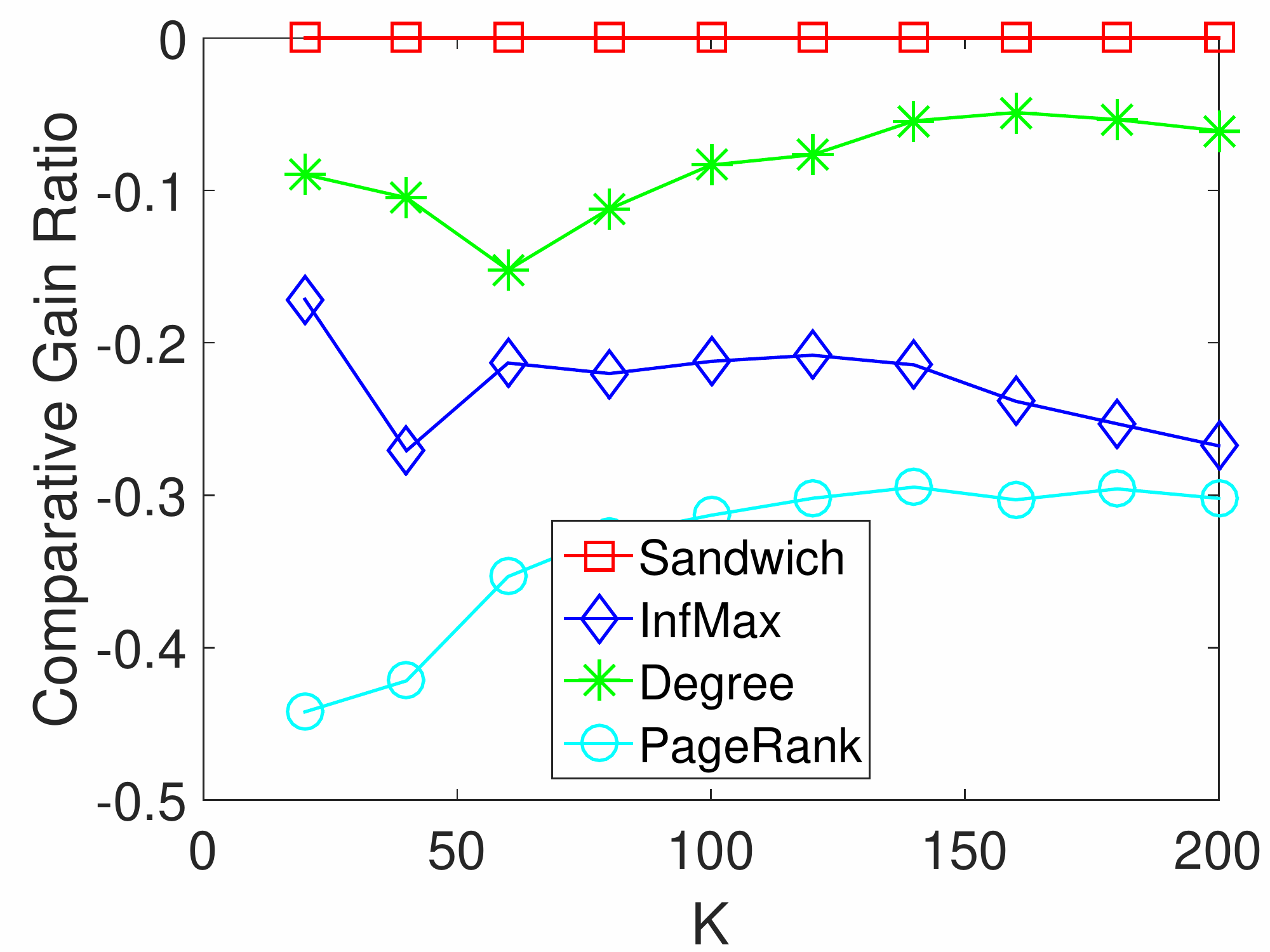}}
    \subfigure[LiveJournal-LT-diffusion]{\includegraphics[width=43mm]{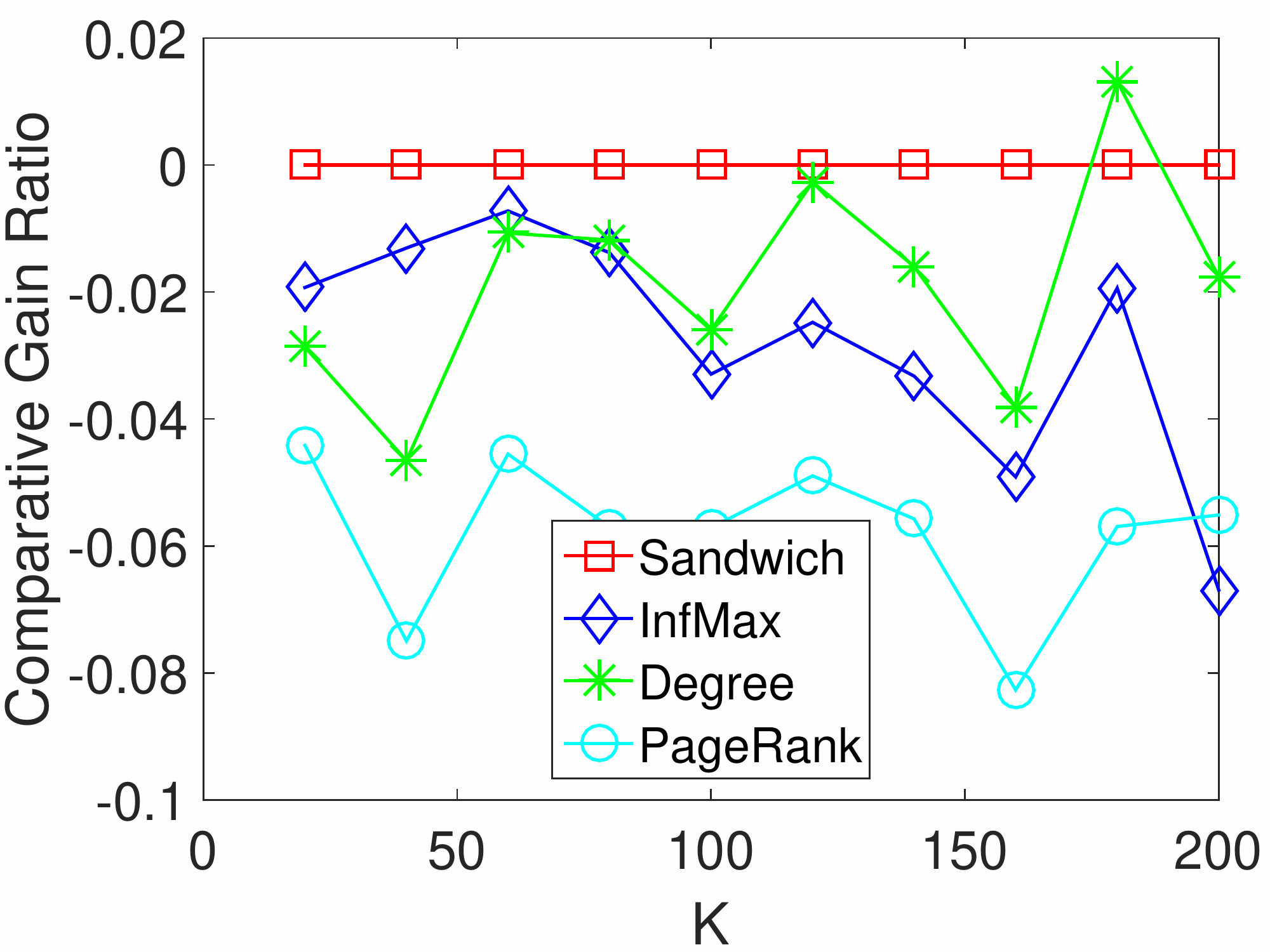}}
    \caption{Information activity on three data sets}
    \label{fig:ia}
\end{figure*}

Fig.~\ref{fig:ia} shows the activity computed by each algorithm on the three data sets, respectively. For better illustration, we report the \emph{comparative gain ratio} instead of the absolute activity value. The comparative gain ratio of an algorithm $\mathcal{A}$ is defined as $\frac{\delta_A(S_{\mathcal{A}})-\delta_A(S)}{\delta_A(S)}$, where $S_{\mathcal{A}}$ and $S$ are the seed sets returned by algorithm $\mathcal{A}$ and the Sandwich algorithm, respectively.

Our algorithm Sandwich almost always has the best performance. Only in very few cases, Sandwich is outperformed marginally. In the uniform settings, algorithm Degree performs well under the IC model but has a relatively bad performance under the LT model. InfMax and PageRank often have a bad performance under both the IC model and the LT model in the uniform settings. In the diffusion settings, InfMax algorithm is a good heuristic under both the IC model and the LT model. Algorithm PageRank performs well on the DBLP data set but has a bad performance on the other two data sets. Algorithm Degree often has a bad performance under both the IC model and the LT model in the diffusion settings. These baseline algorithms only use the properties of the social network or the diffusion process but totally ignore the activity strengths on edges. In contrast, our algorithm utilizes the unbiased estimate of the activity and its lower and upper bounds to solve the problem. This is why our algorithm always has a good performance while the baseline algorithms fail in many cases.

\subsection{Approximation Quality}

A major advantage of our algorithm is that it carries a data dependent approximation ratio. Since the exact approximation is intractable to compute, we report the computable lower bound of the approximation ratio, that is $\frac{(1-\gamma)^2}{(1+\gamma)^2}\cdot(1-e-\epsilon)\cdot\frac{\hat{\delta}_A(S_U)}{\hat{\delta}_U(S_U)}$. Fig.~\ref{fig:approx} shows the results on the three data sets.

The ratio varies in different data sets. On the same data set, the ratios under the IC model and the LT model also differ.  In general, the ratio under the LT model is greater than the one under the IC model in the same activity settings. The ratio does not change much with respect to the size of the seed set $k$. Roughly the ratio increases when $k$ increases. A possible reason is that the gap between the activity and the upper bound shrinks when $k$ increases, since there are more nodes activated with a larger value of $k$.

\begin{figure*}
    \centering
    \subfigure[HepPh]{\includegraphics[width=45mm]{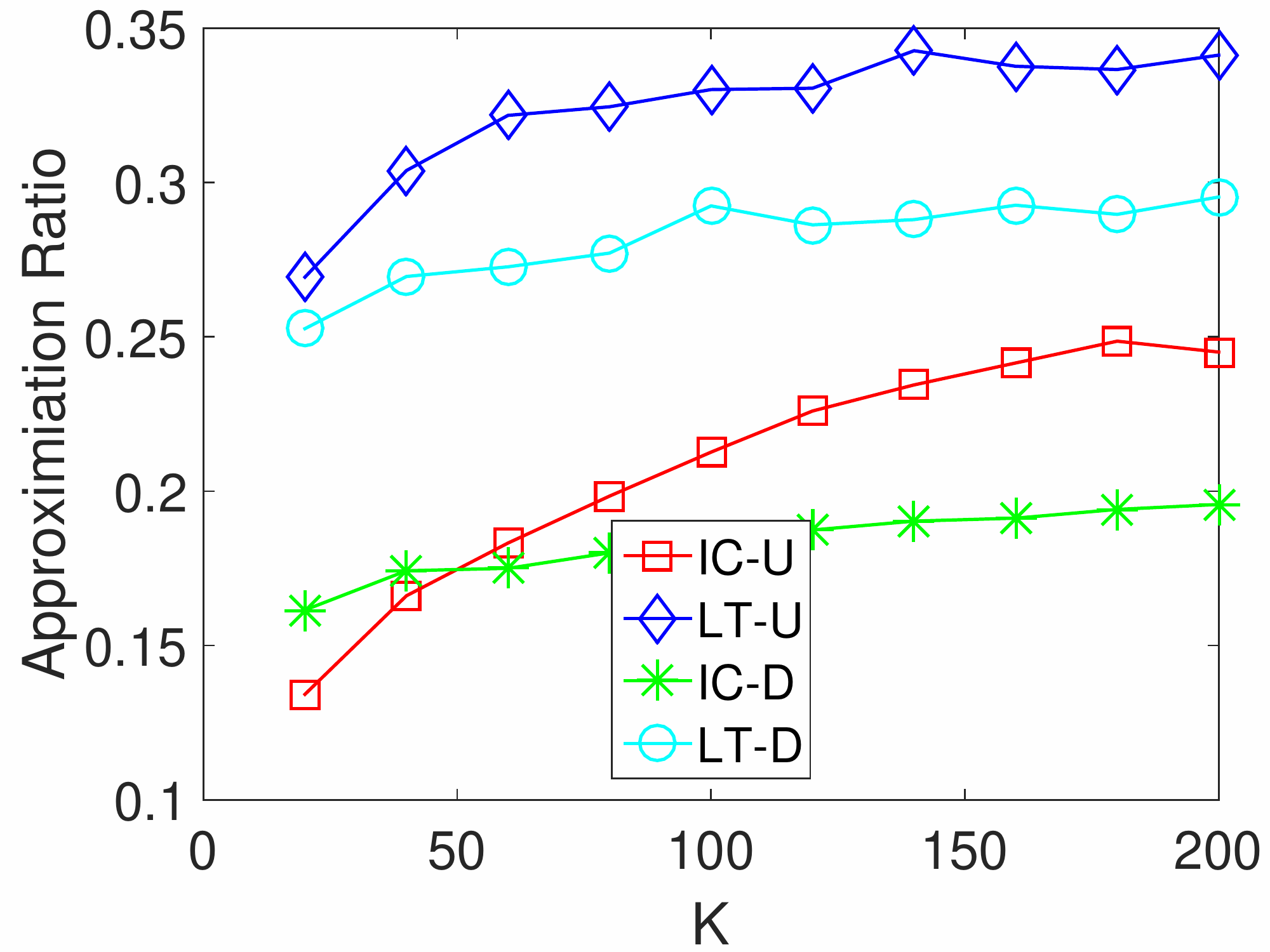}}
    \hspace{0.1cm}
    \subfigure[DBLP]{\includegraphics[width=45mm]{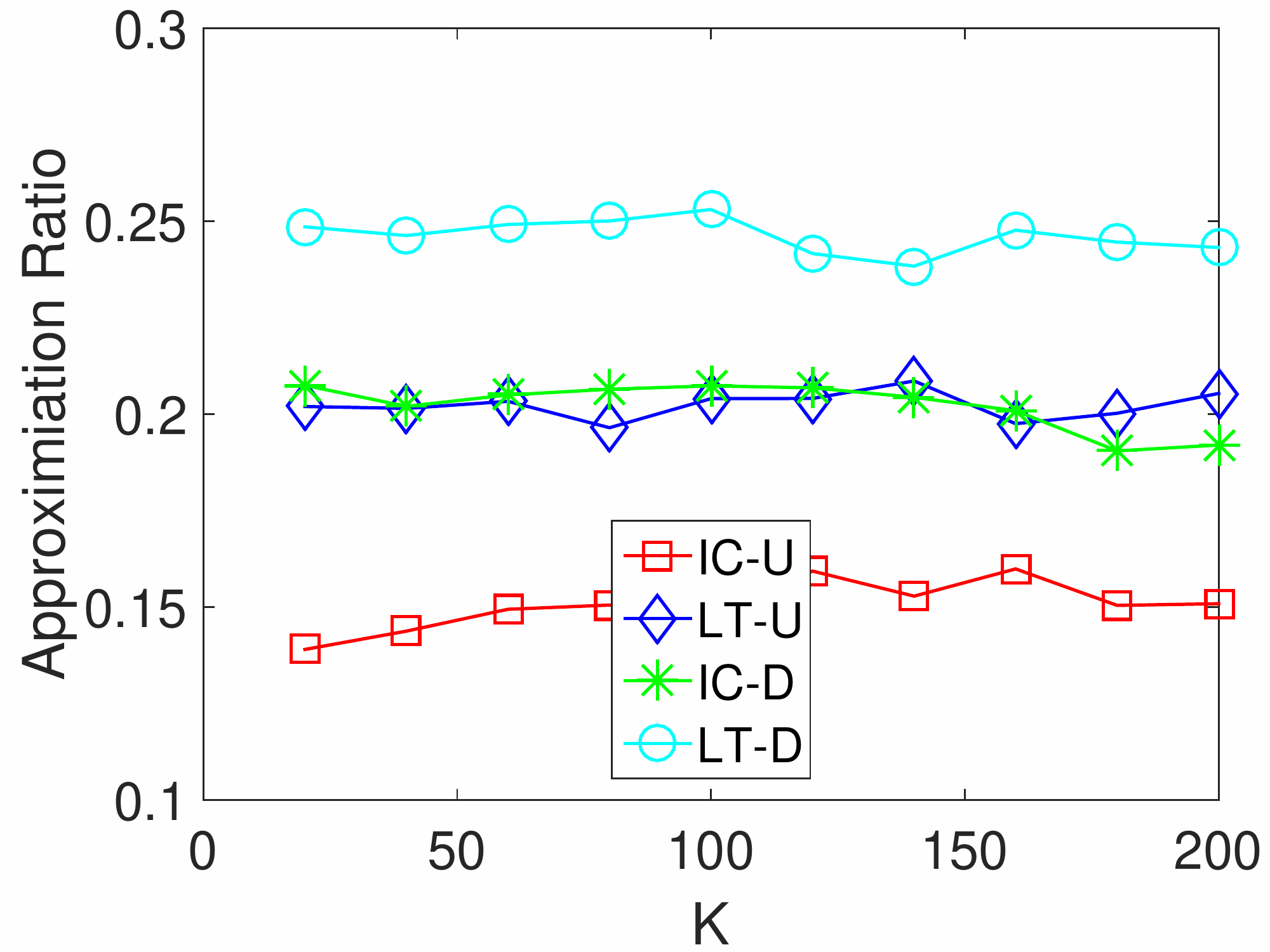}}
    \hspace{0.1cm}
    \subfigure[LiveJournal]{\includegraphics[width=45mm]{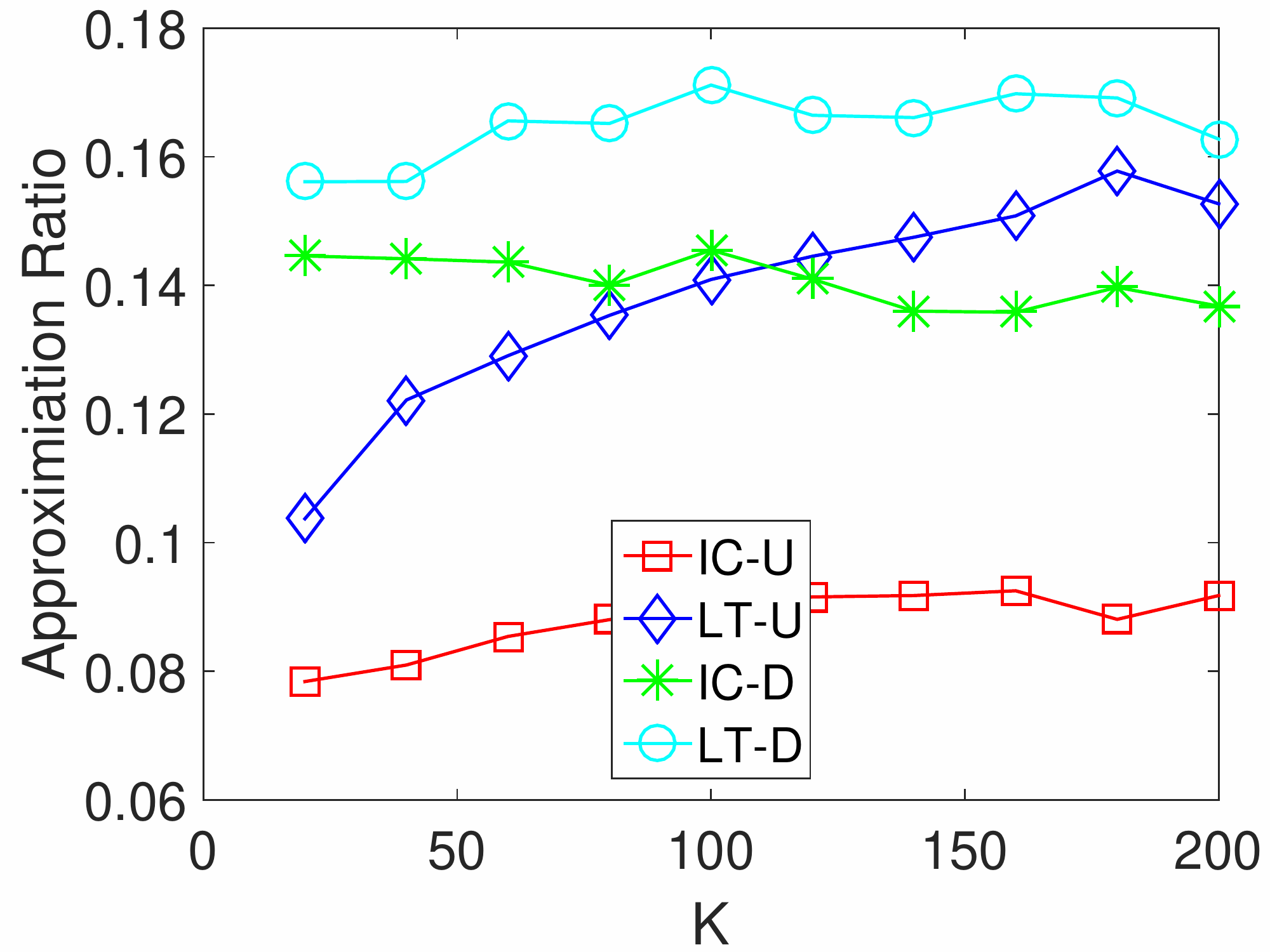}}
    \caption{Approximation ratio on three data sets}
    \label{fig:approx}
\end{figure*}

\subsection{Scalability}

Since the activity settings do not affect the running time, we only report the running time in the uniform case. Fig.~\ref{fig:time} shows the running time on the three data sets.

In most of the cases, the running time of our algorithm decreases when the size of seed set $k$ increases. This is because the time cost in Sandwich depends on the number of sampled hyperedges. According to Lemma~\ref{le:mc}, the expected number of samples is inversely proportional to $\mu_Z$, which is the probability of the event $S\cap R_{g^T}(u)\neq \emptyset \land S\cap R_{g^T}(v)\neq \emptyset$. It increases when $k$ increases. A similar analysis holds for the lower bound and the upper bound. PageRank is faster than our algorithm on the two smaller data sets but slower on the largest data set. Degree and InfMax are more efficient than our algorithm, but they are substantially weaker than ours in effectiveness in many cases. It is worthy noting that our algorithm is actually very efficient. The largest running time is only about $600$ seconds on the largest data set, which has millions of nodes and tens of millions of edges.

\begin{figure}[t]
    \centering
    \subfigure[HepPh-IC]{\includegraphics[width=0.22\textwidth]{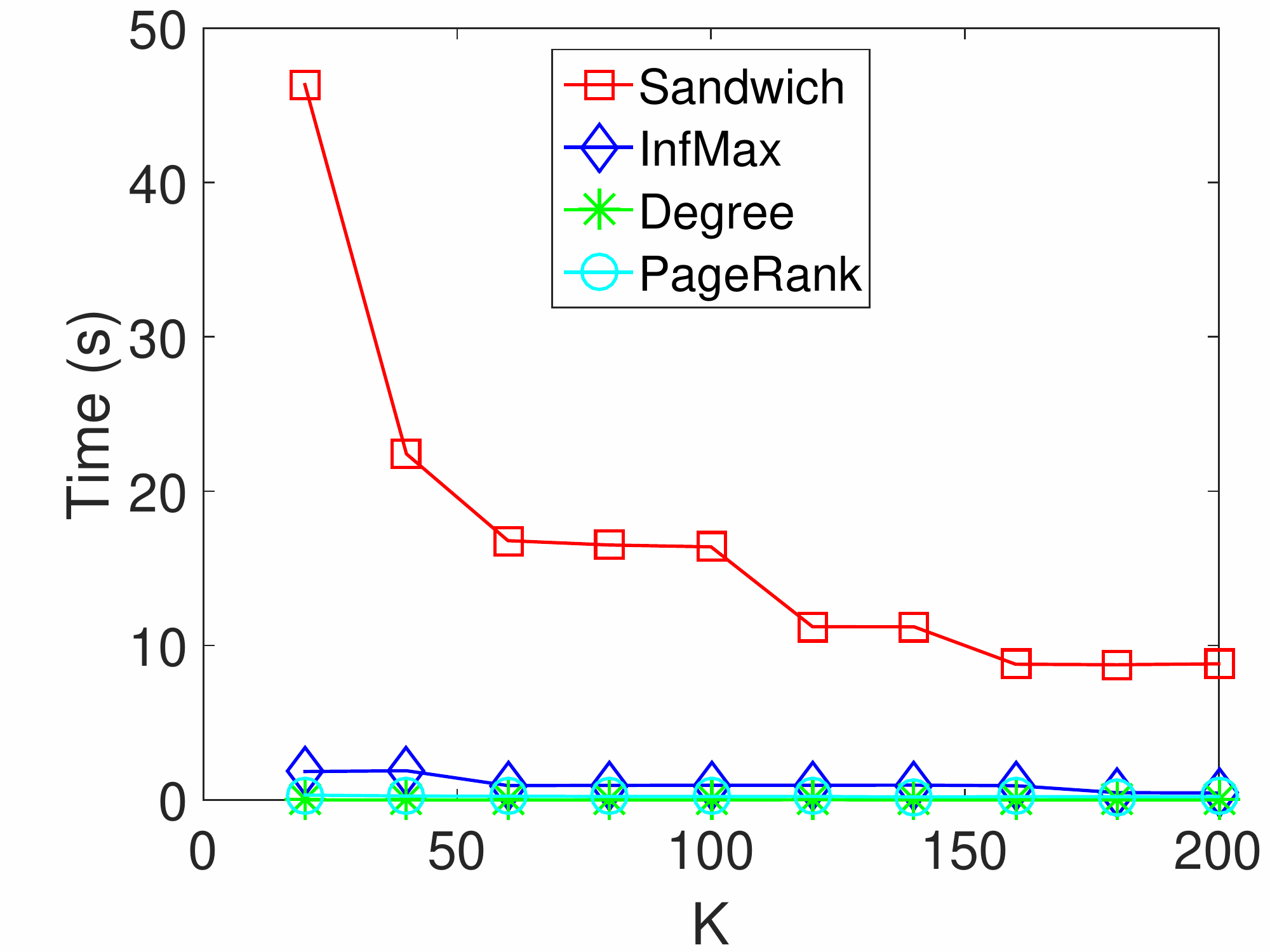}}
    \subfigure[HepPh-LT]{\includegraphics[width=0.22\textwidth]{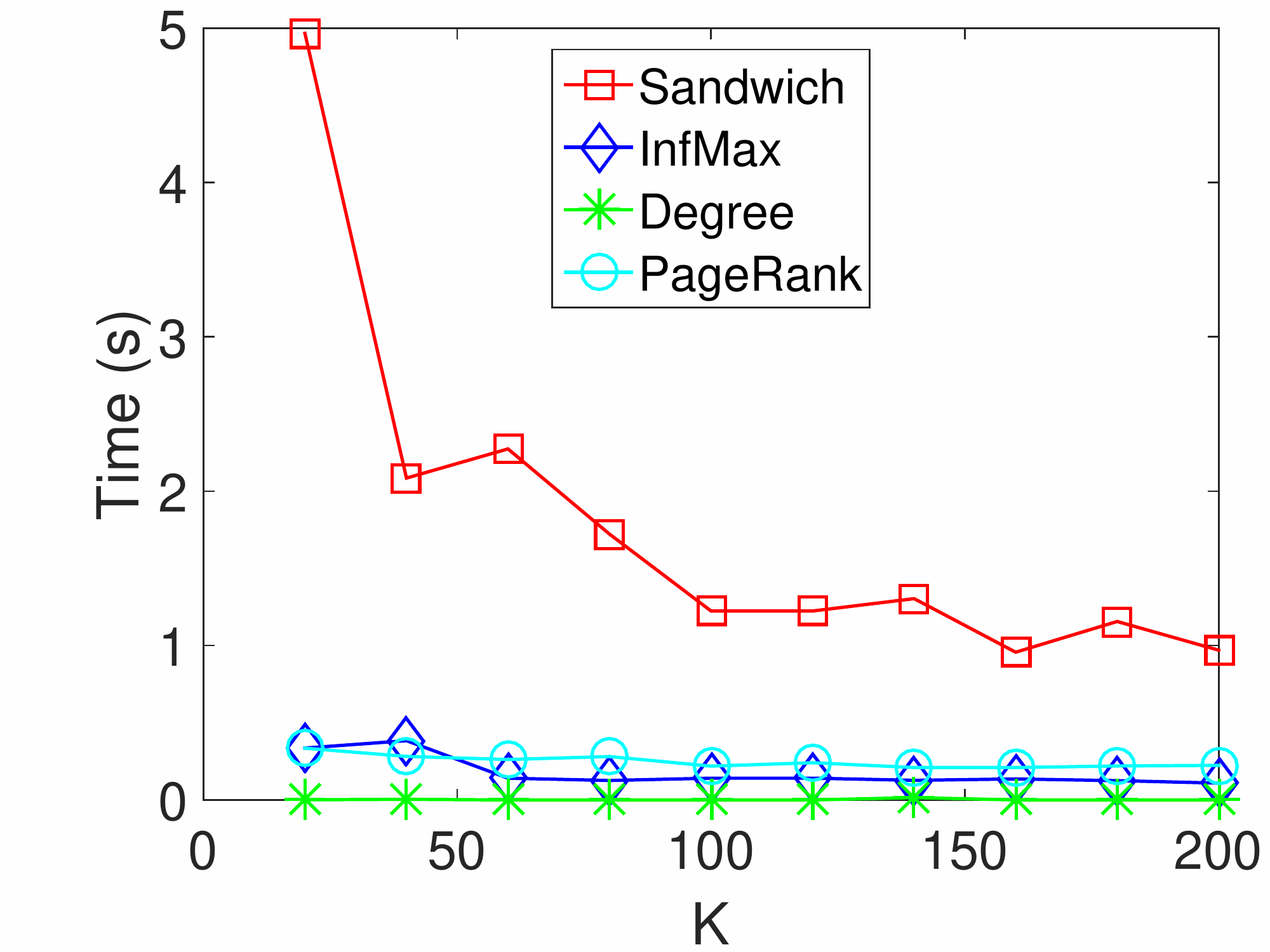}}

    \subfigure[DBLP-IC]{\includegraphics[width=0.22\textwidth]{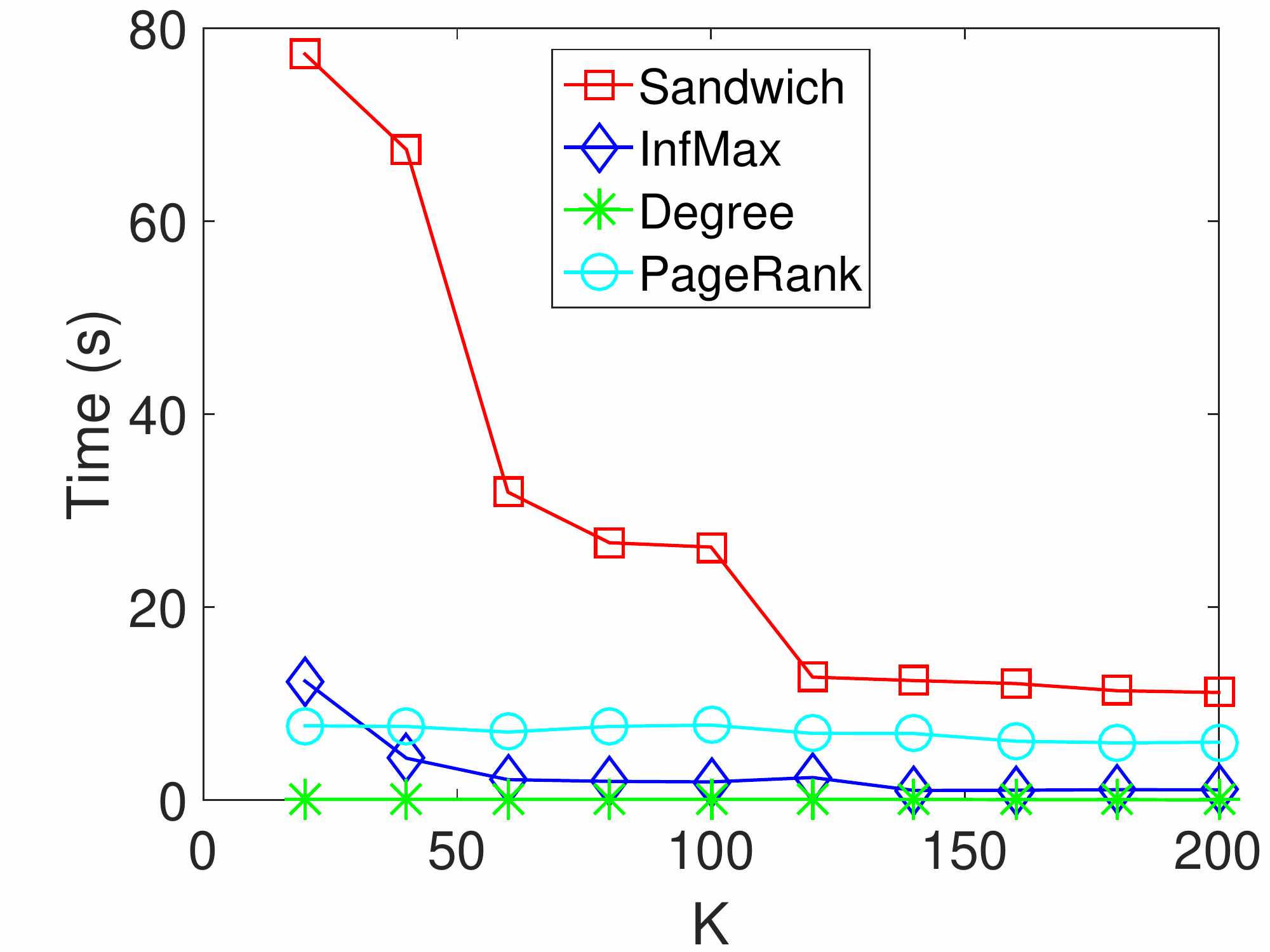}}
    \subfigure[DBLP-LT]{\includegraphics[width=0.22\textwidth]{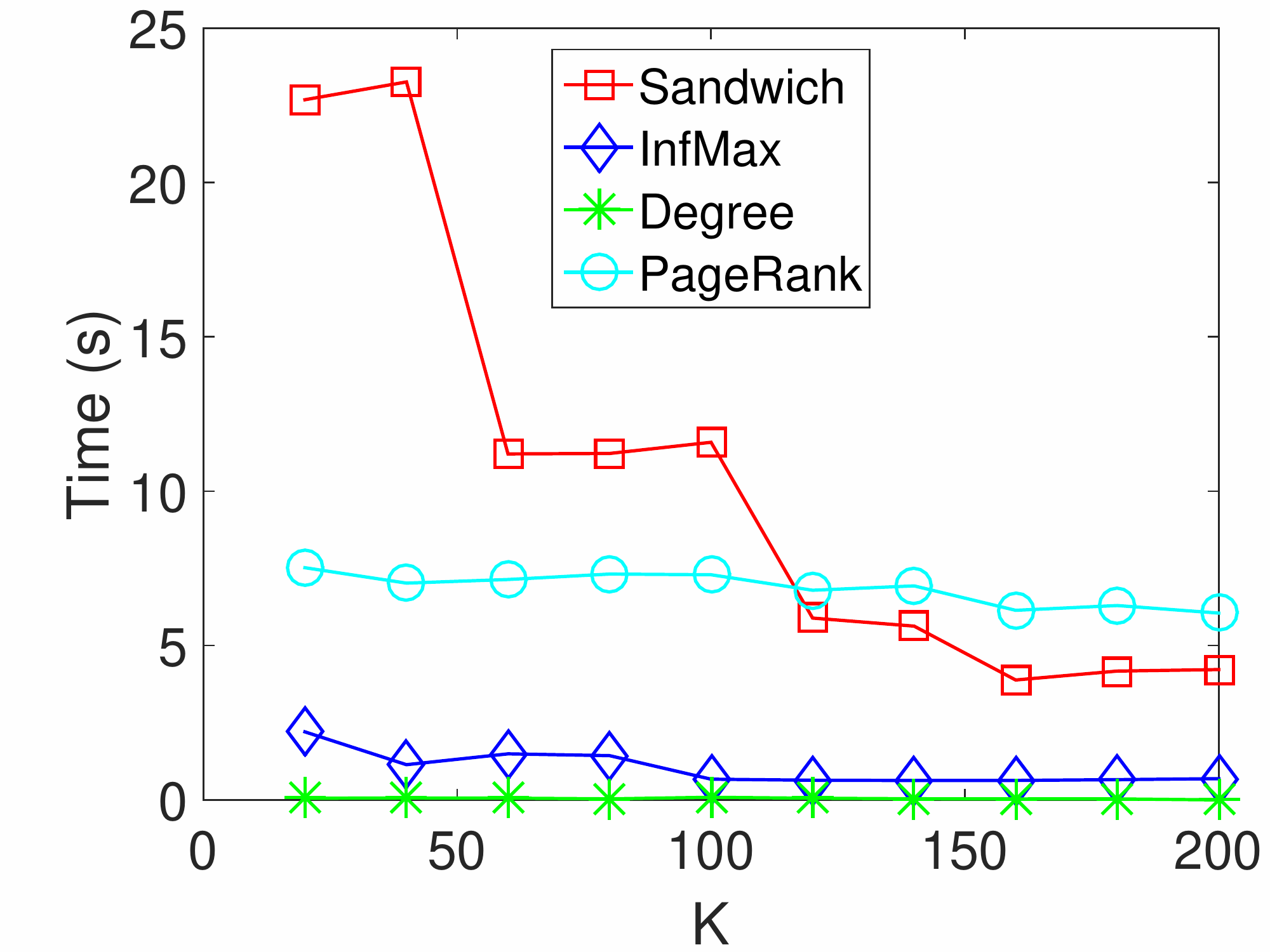}}

    \subfigure[LiveJournal-IC]{\includegraphics[width=0.22\textwidth]{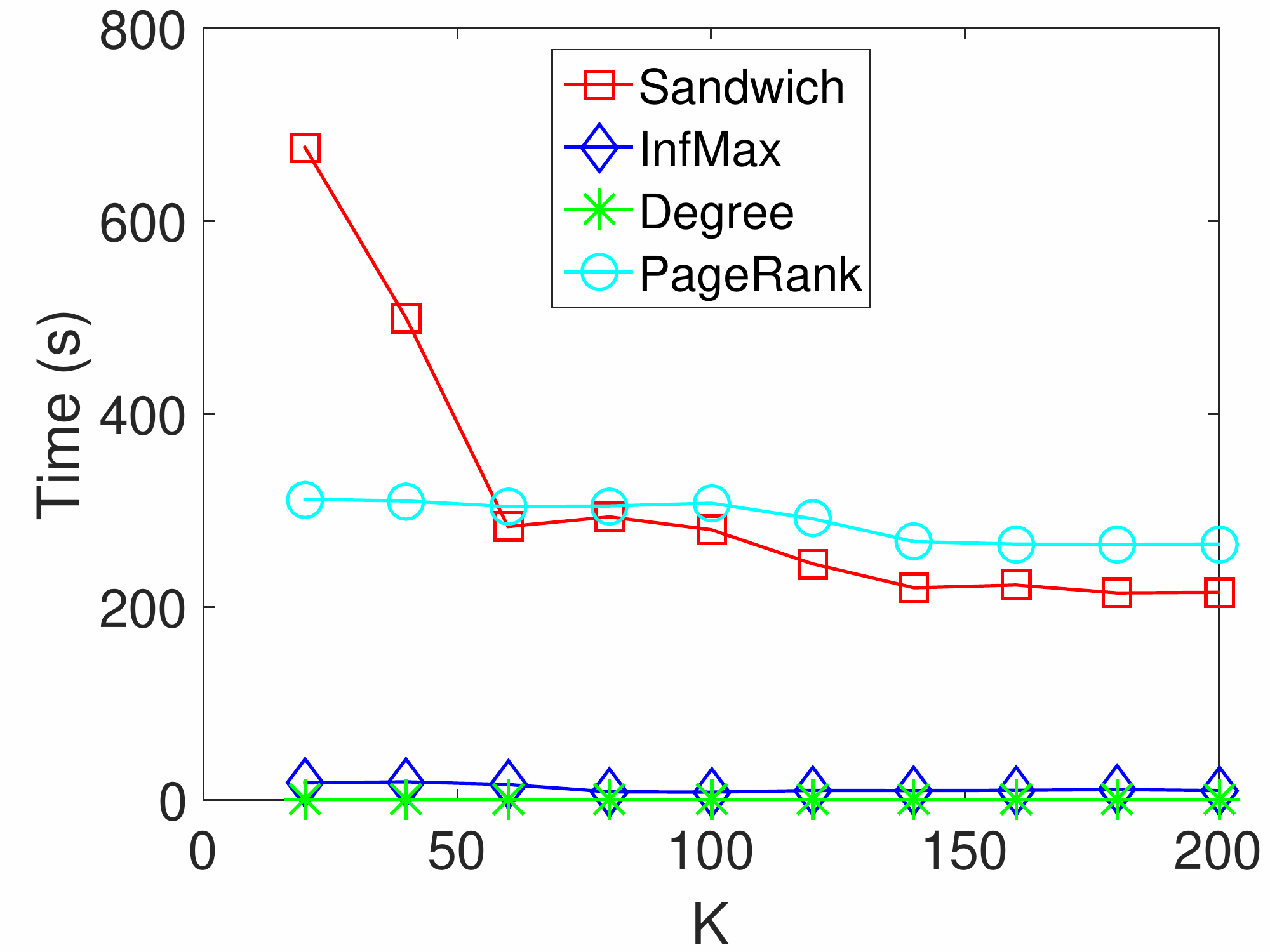}}
    \subfigure[LiveJournal-LT]{\includegraphics[width=0.22\textwidth]{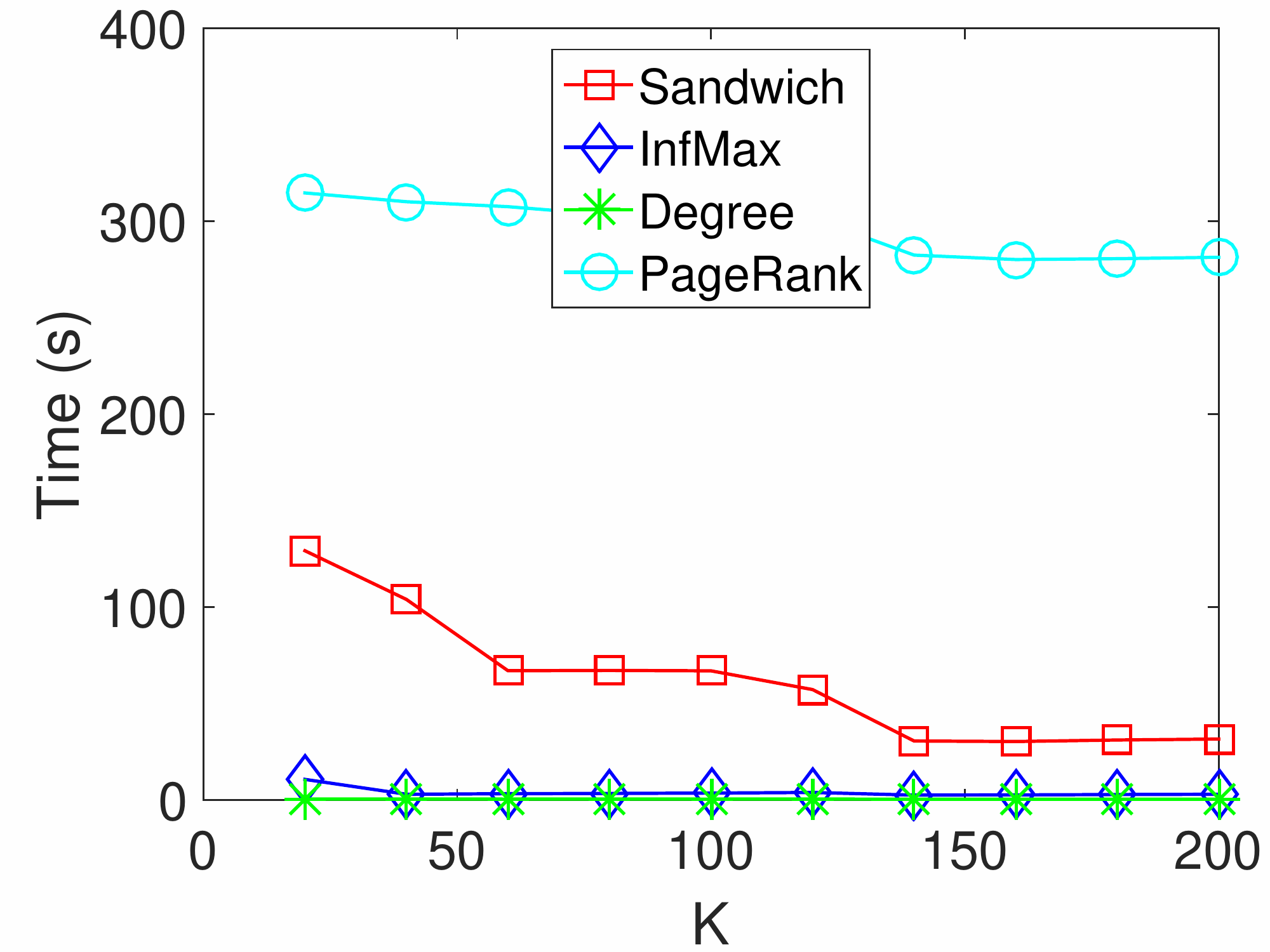}}
    \caption{Running time on three data sets}
    \label{fig:time}
\end{figure}

\subsection{Influence Spread versus Activity}

To explore the relation between influence spread and activity, we report their values in the uniform settings. We choose the uniform settings for our experiments here because in such a situation, the activity is exactly the number of edges between the active nodes. In the diffusion settings, there is no such correspondence. We also calculate their ratio, which is the influence spread against the information activity. Tab.~\ref{tab:case} shows the results on the three data sets.

The ratio differs under different models. In general, the ratio under the LT model is greater than the one under the IC model. Possibly active nodes are more closely connected to each other under the LT model. Interestingly, the influence spread in the DBLP data set is greater than the one in the HepPh data set while the activity in the DBLP data set is smaller than the one in the HepPh data set.
It is probably because the average degree of HepPh is greater than that of DBLP. As a result, there are more edges among the active nodes in the HepPh data set. The results suggest that the ratio is related to the average degree.

We also notice that the ratio is similar when $k=20$ and $k=200$. This result suggests that the relation between the influence spread and the activity does not vary much with respect to the size of seed set. The ratio can be viewed as the average degree of the propagation induced subgraph.

The average degree of the propagation induced subgraph is smaller than the average degree of the whole graph. This is because only a small proportion of the nodes can be activated. Thus, there are many edges between active nodes and inactive nodes. The average degree of the propagation induced subgraph only considers the edges between active nodes. Thus, we report the interaction ratio of the active nodes, which is the number of edges whose both endpoints are active against the number of edges that have at least one active endpoint. The results are shown in Fig.~\ref{fig:bar}. The interaction ratios are not high on all three data sets. This indicates that only a small proportion of the neighbors are activated and interact with the active nodes.  This result demonstrates an essential difference between activity maximization and influence maximization.

\begin{figure}[t]
    \subfigure[$k=20$]{\includegraphics[width=0.22\textwidth]{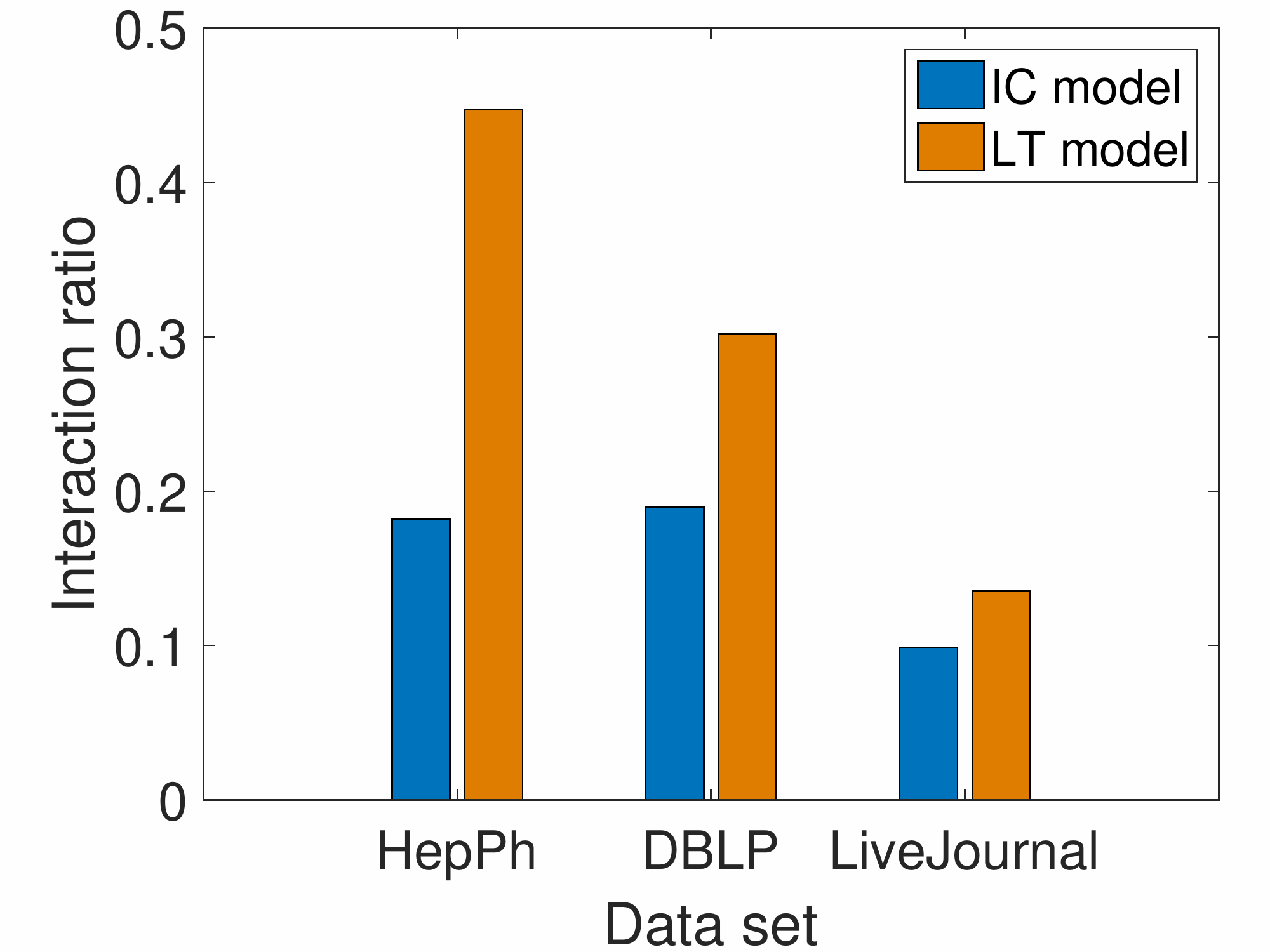}}
    \subfigure[$k=200$]{\includegraphics[width=0.22\textwidth]{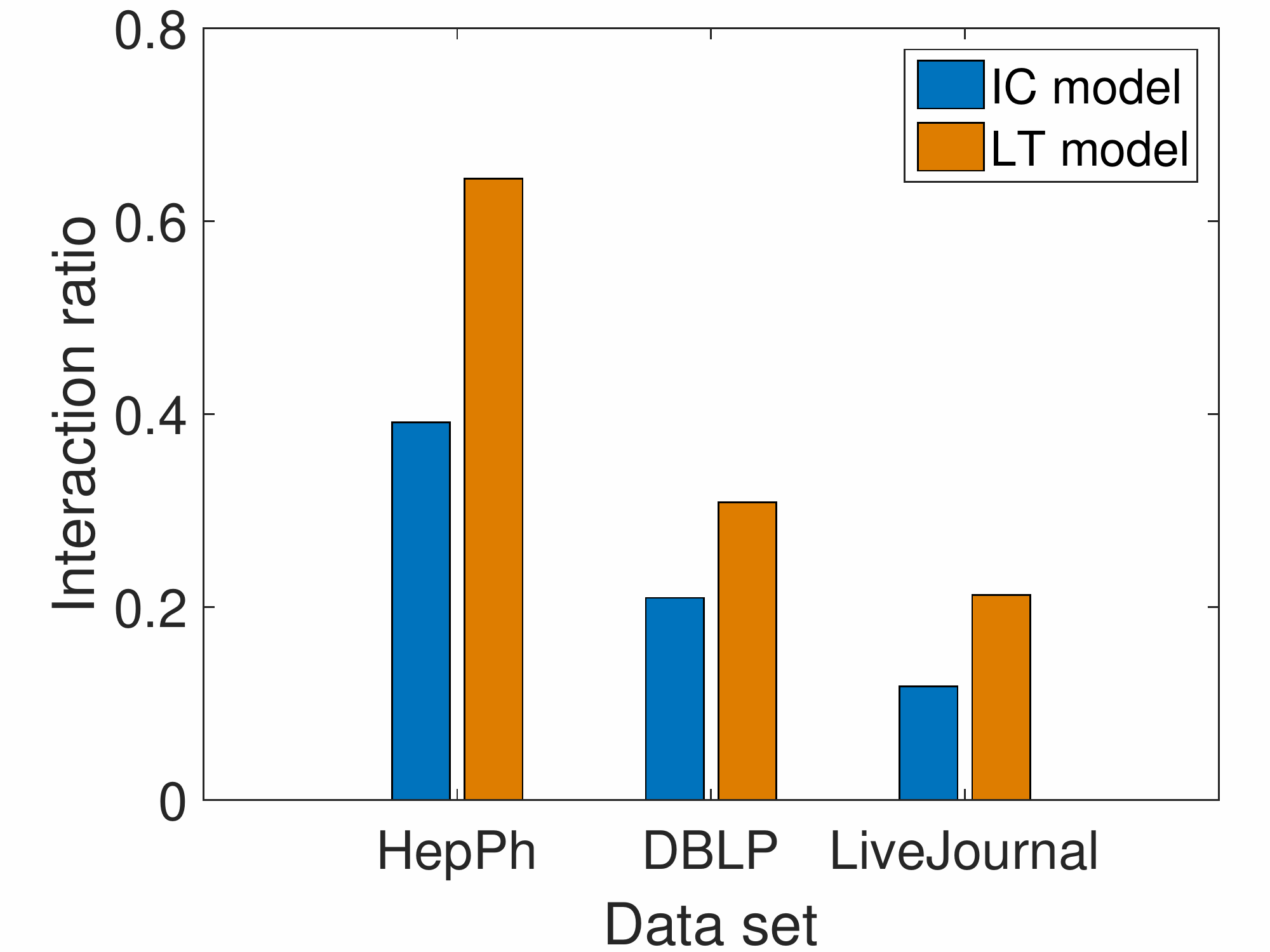}}
    \caption{Interaction ratio on three data sets}
    \label{fig:bar}
\end{figure}

\begin{table*}
    \centering
    \begin{tabular}{c|c|c|c|c|c|c|c|c|c|c|c|c}
    \hline
         \multirow{3}{*}{Data}& \multicolumn{6}{c|}{IC model}& \multicolumn{6}{c}{LT model}\\ \cline{2-13}
         &\multicolumn{3}{c|}{k=20} & \multicolumn{3}{c|}{k=200}& \multicolumn{3}{c|}{k=20} & \multicolumn{3}{c}{k=200}\\ \cline{2-13}
         & influence& activity& ratio& influence& activity& ratio& influence& activity& ratio& influence& activity& ratio\\ \hline

         HepPh& 932&4,868& 5.22& 3,217& 17,016& 5.28& 1,988& 26,017& 13.08& 5,467& 61,524& 11.25 \\
         DBLP& 2,291&3,509& 1.53& 13,764& 21,165& 1.54& 2,834& 5,179& 1.83& 17,445& 32,712& 1.88 \\
         LiveJ& 66,615& 958,95& 1.44& 186,726& 333,598& 1.79& 89,559& 184,842& 2.06& 297,014& 839,334& 2.83\\ \hline
    \end{tabular}
    \caption{Influence spread and information activity on three data sets}
    \label{tab:case}
\end{table*}

\section{Conclusions}\label{sec:con}
In this paper, to address the demand raised in several interesting applications, we proposed and formulated a novel problem, activity maximization. We proved the hardness of the problem under both the IC model and the LT model. We also developed a lower bound and an upper bound of the objective function, and observed several useful properties of the lower bound and the upper bound. We designed a polling based algorithm to solve the problem that carries a data dependent approximation ratio. Our experimental results on three real data sets verified the effectiveness and efficiency of our method. As future work we are interested in learning the activity of user pairs from real-world data.
\bibliographystyle{abbrv}
\bibliography{iam}

\end{sloppy}
\end{document}